\documentclass[%
reprint, 
superscriptaddress,
nofootinbib,
nobibnotes,
 amsmath,amssymb,
 aps, physrev,
]{revtex4-2}

\usepackage{graphicx}
\usepackage{dcolumn}
\usepackage{bm}
\usepackage{hyperref}
\usepackage{xcolor}
\usepackage{comment}
\usepackage{amsmath}

\newcommand {\edit}[1] {{\color{black} #1}}

\begin{document}

\preprint{APS/123-QED}

\title{\edit{Pathogen diversity emerging from coevolutionary dynamics in interconnected systems}}

\author{Davide Zanchetta}
\affiliation{Dipartimento di Fisica e Astronomia “Galileo Galilei”, Universit\`a degli studi di Padova, via Marzolo 8, 35131 Padova, Italy}

\author{Vittoria Bettio}
\affiliation{Dipartimento di Fisica e Astronomia “Galileo Galilei”, Universit\`a degli studi di Padova, via Marzolo 8, 35131 Padova, Italy}

\author{Sandro Azaele}
\altaffiliation{These authors contributed equally to this work.}
\affiliation{Dipartimento di Fisica e Astronomia “Galileo Galilei”, Universit\`a degli studi di Padova, via Marzolo 8, 35131 Padova, Italy}
\affiliation{INFN, Sezione di Padova, via Marzolo 8, Padova, Italy - 35131}
\affiliation{National Biodiversity Future Center, Piazza Marina 61, 90133 Palermo, Italy}

\author{Manlio De Domenico}
\altaffiliation{These authors contributed equally to this work.}
\affiliation{Dipartimento di Fisica e Astronomia “Galileo Galilei”, Universit\`a degli studi di Padova, via Marzolo 8, 35131 Padova, Italy}
\affiliation{INFN, Sezione di Padova, via Marzolo 8, Padova, Italy - 35131}
\affiliation{Padua Center for Network Medicine, University of Padua, Via F. Marzolo 8, 35131 Padova, Italy}

\date{\today}

\begin{abstract}
\edit{The spread of infectious disease and the evolution of antigenically distinct strains are often modeled separately, despite strong feedbacks mediated by host immune memory and heterogeneous contacts. To tackle this challenging problem, we introduce a coevolutionary framework in which transmission occurs on a metapopulation network while mutational exploration of strain space follows a mutation network. In this multiscale model, cross-immunity is encoded by similarity in the latent diffusion geometry of the strain network, so that nearby strains confer partial immune protection. We first identify an effective critical region that controls the transition between extinction, recurrent outbreak episodes, and long-lived endemic persistence, thus characterizing the resulting strain-turnover dynamics. We then derive a replicator–mutator-like equation for strain composition and an explicit dynamical evolutionary landscape induced by the coupling of mutation and transmission. Finally, allowing host heterogeneity to modulate the local mutation structure, we show that spreading across demes can effectively connect otherwise disconnected components of strain space, increasing long-term endemic diversity while producing a non-monotonic change in overall prevalence. Together, our results isolate minimal mechanisms by which immune-mediated competition and network structure can shape antigenic diversification.}
\end{abstract}

\maketitle

\begingroup
\renewcommand{\thefootnote}{}
\footnotetext{\vspace{0.2mm} \\ Contact authors: \\ \href{mailto:davide.zanchetta@unipd.it}{davide.zanchetta@unipd.it} \\ \href{mailto:manlio.dedomenico@unipd.it}{manlio.dedomenico@unipd.it}}
\addtocounter{footnote}{-1}
\endgroup


Epidemic outbreaks have repeatedly impacted the course of human history. In today’s globally connected world, \edit{emerging and re-emerging pathogens of communicable diseases} pose a growing threat to the stability of demographic, economic, and healthcare systems~\edit{\cite{ponton2000emerging,skovgaard2007new,morens2004challenge}}. \edit{An emblematic example is the COVID-19 pandemic, which} underscored the importance of predictive modeling in informing public health responses, \edit{despite the difficulties related to the highly mutational nature of the SARS-CoV-2 virus~\cite{li2021emergence,tao2021biological,lauring2021genetic,hacisuleyman2021vaccine,lauring2021genetic,harvey2021sars}}. Recent estimates suggest that large-scale pandemics may occur roughly once every 50--60 years, with risk projected to increase due to ecological disruption and global mobility \cite{marani2021probability}. Beyond the emergence of novel pathogens, many infectious diseases exhibit endemic dynamics with episodic flare-ups. These recurrent outbreaks often result from a combination of seasonal forcing and pathogen evolution, particularly in viral populations such as influenza \cite{bedford2015global}. Although less disruptive than pandemics, endemic pathogens exert a sustained burden on public health systems and disproportionately affect vulnerable populations \cite{lemey2020accommodating}.

Mathematical models of epidemics~\cite{hethcote2000mathematics} have produced much insight into outbreak dynamics, for example concerning epidemic thresholds, or lack thereof~\cite{PastorSatorras2001,pastor2002epidemic} (\edit{see \cite{pastor2015epidemic} for a review}). However, modeling the interplay between mutational diversification and epidemic dynamics remains challenging, also due to the inherent complexity of genetic and antigenic spaces~\cite{grenfell2004unifying,mora2023quantitative,lamata2025genotype}. Even modest genome lengths $L$ produce an exponentially large number of possible sequences (\(=\left(\text{number of bases}\right)^L\)), but the relevant space of viable and immunologically distinct strains most likely forms a smaller, irregularly connected network~\cite{koelle2006epochal}, since not all possible sequences are biologically relevant.  \edit{This fact allows one to employ sophisticated inferential techniques that can efficiently explore the relevant evolutionary landscape~\cite{lee2018deep,starr2020deep,dadonaite2024spike}}. \edit{However,} the available mutational pathways might change according to physiological differences between different hosts~\cite{ekroth2021host,jones2021viral,allman2022heterogeneity}. Previous works have sought to attack this problem by homogeneous one- or two-dimensional descriptions of strain space~\cite{gog2002dynamics,marchi2021antigenic,chardes2023evolutionary}, which is motivated by the observation that flu strains follow mostly linear trajectories in latent spaces described using appropriate techniques~\cite{smith2004mapping}. \edit{There is evidence that this also holds for SARS-CoV-2~\cite{wilks2023mapping}. At the same time, it has been suggested that the apparent low-dimensionality of antigenic spaces is an emergent -- rather than fundamental -- feature~\cite{moore2018high}}. While the temporal patterns described by the aforementioned models match our broad expectations of epidemic trajectories, model dynamics should generally also allow the resurgence of previously excluded strains, which is a well-documented occurrence~\cite{phadungsombat2018emergence,sharma2014dominance,paiva2013evolutionary,langat2017genome,zintgraff2025evolving,leslie2004hiv,chopera2008transmission}. 

\begin{figure*}
    \centering
    \includegraphics[width=\textwidth]{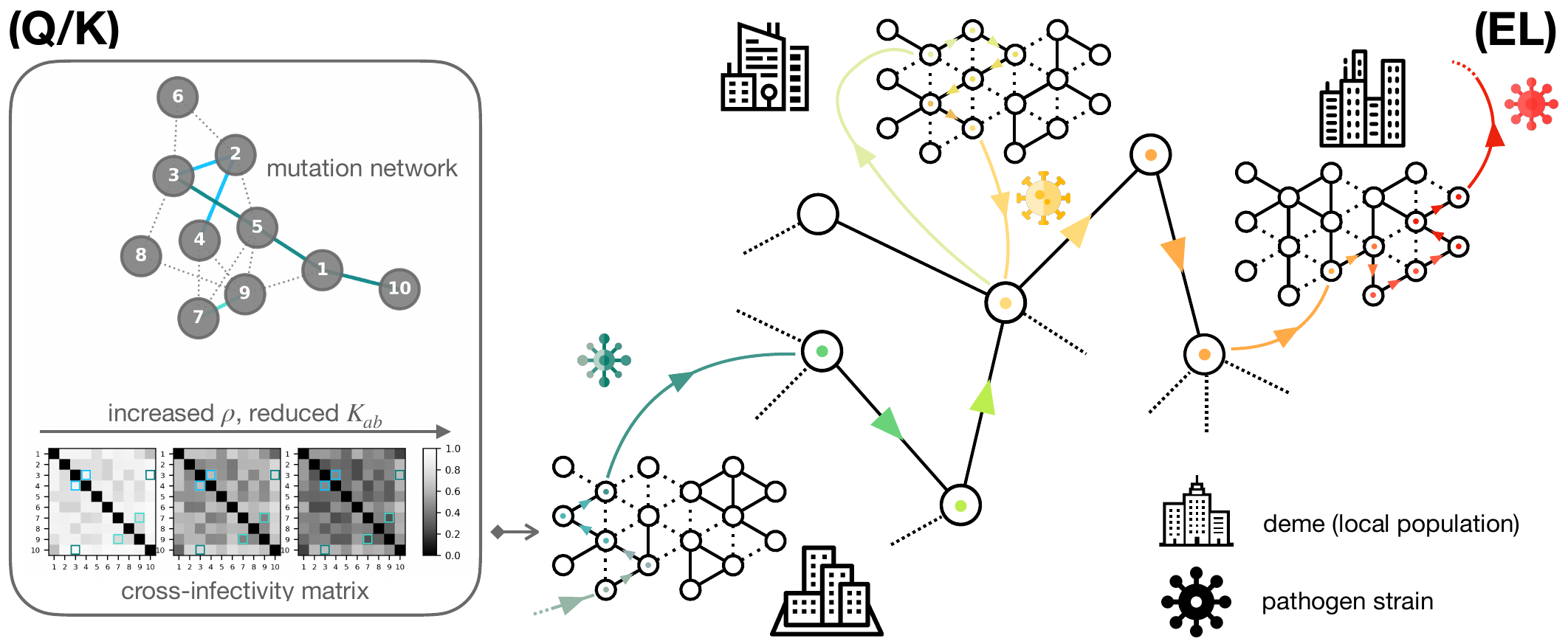} 
    \caption{The evo-SIS framework unfolds over two structures. Locally \textbf{(Q/K)}, pathogen strains change according to a mutation network, which also defines cross-infectivity according to the strength of cross-immunity. Globally \textbf{(EL)}, the mutation and metapopulation structures become intertwined, as the pathogen spreads across heterogeneous demes and an effective evolutionary landscape emerges.}
    \label{fig:1}
\end{figure*}

To the best of our knowledge, the existing literature lacks a framework for modeling pathogen evolution in epidemic dynamics \edit{unfolding on a complex social networks}, which describes strains as units in a complex heterogeneous space, and acknowledges the \edit{interplay} of this \edit{multiscale} heterogeneity \edit{with} cross-immunity effects. To bridge this theoretical gap, here we develop an evolutionary susceptible-infected-susceptible (evo-SIS) framework to couple pathogen spread and evolution across two heterogeneous structures \edit{existing on distinct scales}: a metapopulation network and a mutation network (Fig.~\ref{fig:1}). We model mutations as a diffusion process~\cite{gog2002dynamics} and, correspondingly, we map the mutation network into a cross-infectivity matrix by \edit{its latent} diffusion distance~\cite{de2017diffusion,beretta2025latent}, thus encoding cross-immunity effects. \edit{Crucially, this makes the framework suitable for exploring vaccination strategies and to assess preparedness against future pandemics~\cite{syrowatka2021leveraging,klamser2023enhancing,kraemer2025artificial,castioni2024rebound}, thus complementing empirical studies~\cite{jiang2026learning}.}

Our model comprises only the fundamental processes of infection, recovery and mutations, representing a coarse approximation of reality. In fact, important aspects such as multiple immune memories have been modeled independently with some level of realism, under simplifying assumptions~\cite{blot2025host}. In this work, instead, we focus on the heterogeneous multiscale evolutionary-epidemiological structure.
Here, different pathogen strains compete by inducing mutually detrimental immune protection in previously infected hosts; this indirect interaction is especially strong between antigenically similar strains~\cite{pilosof2019competition}. From an ecological point of view, the hosts' population is an environment for the pathogen strains' community, and immune response is an environmental response to multi-strain population dynamics, unfolding on some characteristic time scales~\cite{ferraro2025synchronization,lion2018theoretical}. While the co-evolution of strains community and environment implies an overlap of biological rates, in specific regimes one may assume a separation of time scales. It is thus possible to recast ecological, population, and evolutionary dynamics as the study of adaptive landscapes~\cite{svensson2012adaptive,catalan2017adaptive}.

\edit{In the remainder of this work, we show that coupling mutation and immune-mediated competition reshapes epidemic outcomes within a minimal evo-SIS framework. We first identify an effective critical region controlling the transition between extinction, recurrent outbreak episodes and long-lived endemic persistence. We then show that near-criticality naturally generates SIR-like boom–bust episodes coexisting with sustained circulation and rapid strain turnover. Using an adiabatic (quasi-stationary and beyond) elimination of susceptibles near endemic steady states, we derive an effective replicator–mutator description for strain composition -- connecting our model with a well-established theoretical framework~\cite{komarova2004replicator} -- and an explicit dynamical evolutionary landscape induced by the coupling of mutation and transmission. Finally, we show that host heterogeneity intertwines contact and mutation structures to open extended evolutionary paths, effectively connecting otherwise disconnected components of strain space, boosting long-term endemic diversity, and producing a non-monotonic dependence of prevalence on heterogeneity.}

\section*{Model}

\edit{In the following, we consider a metapopulation network $A$ with $N$ demes and adjacency matrix $\mathbf{A}$. Such demes might represent populations in different patches of the geographic area of interest, such as countries (or sub-national regions) within the whole planet or sub-national regions within a country. The appropriate choice depends on the application, since some pathogens are limited to specific areas because of favorable environmental conditions, whereas others can spread widely.
For simplicity, we approximate stochastic within-deme contacts as well mixed.
}

\edit{Similarly, we consider a mutation network $Q$ with $M$ strains and adjacency matrix $\mathbf{Q}$, assumed to be representative of the pathogen's mutational dynamics across individuals and demes.}

\edit{To avoid confusion between indices, we will use letters $x,y,z\dots$ to mark deme indices, and $a,b,c\dots$ to mark strain indices.}


We adopt a Force of Infection (FoI) description of spreading, that is without explicit host mobility, where contacts -- and thus possible infections between nodes -- take place at a baseline rate $\beta$ according to the network \(\mathbf{A}\)\edit{; the details entailed in this choice will become clear later, when we give the full form of dynamics.} Infected nodes recover at rate $\gamma$ into a susceptible class which carries the same strain index as the previous infection. Mutations, which track the dominant strain of each node, take place at rate $\eta$, and are described as diffusion on the mutation network, regulated by the Laplacian matrix \(\mathbf L_Q\), \((L_Q)_{ab} = \delta_{ab}\left(\sum_c Q_{cb}\right) - Q_{ab}\). This effective structure is related to, but distinct from, the network of possible mutations of individual pathogen organisms. For now, we simply point out that labeling infected nodes with the dominant strain at each time is a valid assumption when within-host \edit{(and hence within-deme)} diversity is low. This is accurate if the dominant strain turnover is rapid, as discussed in detail in Appendix~\ref{app:RM_within_host}.

Both networks are assumed to be undirected and unweighted. The ``undirectness'' of the metapopulation network is easily justified: contact events between individuals normally entail the reciprocal possibility of infection. We generally assume that mutations between strains, at the effective level described above, are equally likely in both directions. However, to incorporate effects leading to sequential strain turnover, we also consider directed mutation networks, as discussed in Appendix~\ref{app:directed_mutations}. We leave the analysis of pathogen-level mutational irreversibility to future works.

Empirically, it is known that immune memory of a certain strain also protects a host from infections by similar strains, a phenomenon known as \textit{cross-immunity}~\cite{andreasen1997dynamics}. Correspondingly, in our model, the baseline infective rate \(\beta\) is modulated by the \textit{cross-infectivity} matrix \(\mathbf{K}\), which we define to be the diffusion distance matrix~\cite{de2017diffusion,beretta2025latent} associated to \(\mathbf{Q}\): \(\mathbf K =\mathbf d_\rho(\mathbf Q)\) (see Appendix~\ref{app:diffusion_distance} for details). Here, the dimensionless parameter $\rho \geq 0$ is a phenomenological quantifier of cross-immunity. For $\rho = 0$, $K_{ab} \propto  1 - \delta_{ab} $, and immunity is restricted to the last strain encountered: susceptibles of strain $a$ can be infected by strain $b$ at the full rate $\beta$. As $\rho$ increases, most entries of \(\mathbf K\) will decrease, describing broad cross-immunity. Generally, infection success depends on how far apart strains are in mutation space, and not just whether they are identical.

Starting from the microscopic (i.e. individual level) stochastic dynamics, we derive node-level evolution equations for the marginal states. These equations correspond to a first-order mean-field approximation, where correlations between the states of different individuals are neglected. The state of each node \( x \) at time \(t\) is \(\Xi_x(t)=\left\{I_{xa}(t),S_{xa}(t)\right\}_{a=1}^M\). Each node represents a large and well-mixed sub-population, i.e. a \textit{deme}, in a meta-population. We interpret the state \(\Xi_x(t)\) as the composition of deme \(x\) in a mean-field sense, i.e. in terms of fractions of population: \(I_{xa}\) (\(S_{xa}\)) represents the fraction of deme \(x\) infected with (having immune memory of) strain \(a\). The state of each node is subject to the constraint \(\sum_a \left(I_{xa}(t) + S_{xa}(t) \right) = N_x \ \forall t\), with \(\sum_x N_x = N_\text{tot}\), where \(N_x\) (\(N_\text{tot}\)) represents the local (global) population. To limit the number of parameters, and to keep normalization consistent, we consider \(N_x = 1/N \ \forall x\), so that \(N_\text{tot} = 1\). Assuming that populations' sizes are constant means that we will neglect demographic dynamics: the death of each host is balanced by the birth of another host with the same immune memory.

We assume that the rates of infection, recovery and mutation are strain-independent. By a rescaling, we can absorb \(\gamma\) in the other rates, so that $\beta$ and $\eta$ become dimensionless ratios quantifying the typical number of contacts and mutations over the time a host remains infected. Then, the dynamics of the full system, omitting time arguments, is written as

\begin{subequations}\label{eq:model_full}
    \begin{equation}\label{eq:model_full1}
        \dot{I}_{xa} = +\beta \sum_y \sum_b A_{xy}K_{ba}S_{xb}I_{ya} - I_{xa} - \eta \sum_b \left(L_Q\right)_{ab} I_{xb} \\
    \end{equation}
    \begin{equation}\label{eq:model_full2}
        \dot{S}_{xa} = -\beta \sum_y \sum_b A_{xy}K_{ab}S_{xa}I_{yb} + I_{xa}
    \end{equation}  
\end{subequations}

\edit{Note that we use a Force of Infection (FoI) approach to model the spreading process at an effective level~\cite{balcan2009multiscale,belik2011natural,bosetti2020heterogeneity,maniscalco2025critical}. Specifically, we write the deme- and strain-specific FoI as \(\lambda_{xa}(t) = \beta\sum_{y=1}^N\sum_{b=1}^M A_{xy}K_{ab}I_{yb}(t)\), leading to the more compact form for eq.~\eqref{eq:model_full2} 

\begin{equation}\label{eq:model_full_FoI2}
    \dot{S}_{xa} = -\lambda_{xa}S_{xa} + I_{xa}
\end{equation}

We also remark that a similar rewriting is not possible for eq.~\eqref{eq:model_full1}, owing to the asymmetric nature of infective events. Nevertheless, the formulation in terms of FoI is crucial in integrating out the host population and deriving evolutionary dynamics of pathogens, as explained later in the paper.}

Eq.~\eqref{eq:model_full1} describes the change in the fraction of individuals in deme \(x\) infected with strain \(a\). The first term on the r.h.s. quantifies the infection of susceptible individuals (with any immune memory \(b\)) at node \(x\) from infected individuals of strain \(a\) in the neighborhood of \(x\). The second term indicates recovery of infected hosts. The third term denotes mutations, changing the strain of infected individuals within each deme.

Eq.~\eqref{eq:model_full2} instead describes the change in the fraction of susceptible individuals in deme \(x\), with immune memory of strain \(a\). The terms on the r.h.s. of this equations are analogous to those of eq.~\eqref{eq:model_full1}, but for the lack of a mutation term. Note that the cross-infectivity matrix \(\mathbf K\) appears with a different ordering of indices in eqs.~\eqref{eq:model_full1} and~\eqref{eq:model_full2}; in particular, it is transposed in the former equation. While the diffusion distance used to define \(\mathbf K\) is symmetric (see Appendix~\ref{app:diffusion_distance}), the distinction is crucial in defining directed structures, as discussed later in the text and in Appendix~\ref{app:directed_mutations}. \edit{We also remark that, in our framework, the mutation matrix \(\mathbf Q\) belongs to a matrix ensemble. Therefore, both \(\mathbf Q\) and \(\mathbf K\) are random matrices: while the evolution of eq.~\eqref{eq:model_full} is deterministic, its configuration is stochastic.}

\section*{Results}

In this section, we first analyze our model by \edit{considering a single deme, where the core features of dynamics can be better understood}. After outlining the broad phenomenology of the model, we analyze its behavior in the critical region of parameters identified by the epidemic threshold, and compare the model predictions to empirical data. Afterwards, we instead consider the regime of large Forces of Infection (FoI) and perform a systematic expansion to obtain effective dynamics valid in proximity of steady states. We use this to characterize a heterogeneous host population, where a non-trivial metapopulation structure can create extended evolutionary paths, giving rise to a dynamical evolutionary landscape.

\subsection{Dynamics within a single well-mixed deme} 

\edit{We first describe a single, well-mixed population, i.e. characterized by a all-to-all contact structure. Suppressing spatial indices which appear in eq.~\eqref{eq:model_full}}, dynamics read

\begin{subequations}\label{eq:model_homMF}
    \begin{equation}\label{eq:model_homMF1}
        \dot{I}_{a} = +\beta \sum_b K_{ba}S_{b}I_{a} - I_{a} - \eta \sum_b \left(L_Q\right)_{ab} I_{b} \\
    \end{equation}
    \begin{equation}\label{eq:model_homMF2}
        \dot{S}_{a} = -\beta \sum_b K_{ab}S_{a}I_{b} + I_{a}
    \end{equation}  
\end{subequations}

\edit{A commonly used indicator of the evolution of an epidemic is the \textit{epidemic size} (or \textit{total prevalence)}, defined as the total number of infected individual at a given time \(t\), given by \(I(t) = \sum_a I_a(t)\). Moreover, we are interested in describing dynamics of pathogen diversity, as observed through the prevalences of hosts infected with different strains. In our framework, we consider a fixed number \(M\) of strains, whose prevalences, however, can change widely. To quantify diversity, then, it is not sufficient to enumerate strains which are detected in the host population. Rather, any measure of diversity should naturally take into account that diversity patterns are more strongly influenced by the most abundant strains. Therefore,} in the following, we will estimate the effective diversity of \(M\) strains with relative prevalences \(\{p_a\}_{a=1}^M\) using evenness, which is defined as

\begin{equation}\label{eq:evenness}
    E = \frac{1}{M}\exp{\left(-\sum_a p_{a}\log{\left(p_{a}\right)}\right)} \ ,
\end{equation}

The exponent is the Shannon entropy of relative prevalences, by which properties it holds \edit{for large M} \( 0 \leq E \leq 1 \). \edit{\(E\) is close to 0 when \(p_a \sim 1\) for one strain \(a\) and \(p_b \sim 0\) for all other strains \(b \neq a\). Conversely, \(E \sim 1\) if \(p_a \sim 1/M \forall a\)}.
 This quantity has been shown~\cite{zanchetta2025emergence} to upper bound the detection rate of strains under idealized sampling conditions.

\emph{Critical transmission rate.} The analysis of the epidemic threshold \edit{for a seed strain invading an immunologically naive population is analogous to that of} the basic SIS model, \edit{since mutations play no role in this instance}. Consider the system being invaded by strain \(1\), \(I_1(0) = I_0 \ll 1\). \edit{In order to describe a population which has had no prior contact with any strain of the pathogen, we set initial conditions so that all susceptibles carry the index M, and therefore \(S_M(0) = 1 - I_0\), and we disconnect strain \(M\) from the rest of the network, so that \(K_{1M} \equiv \kappa(\rho) \gtrapprox 1 \), \(\forall\rho \geq 0\) (see Appendix~\ref{app:diffusion_distance})}. Then, summing over strain indices to consider total prevalence, the invasion dynamics reads

\begin{equation}
    \left.\dot I / I \right|_{t=0} = \kappa(\rho)\beta  - 1 + \mathcal{O}(I_0) \ ,
\end{equation}

which is exactly the basic result on the epidemic threshold: \edit{the seed strain can increase in prevalence if \(\beta > \beta_{\text{crit}} \equiv 1/\kappa(\rho)\)}. It should be noted that \(\beta > \beta_\text{crit}\) only implies an increase in epidemic size at the beginning of dynamics. In fact, depending on the strength of cross-immunity, this increase can be short lived.

\begin{figure*}
    \centering
    \begin{minipage}{\textwidth}
    \includegraphics[width=0.6\textwidth]{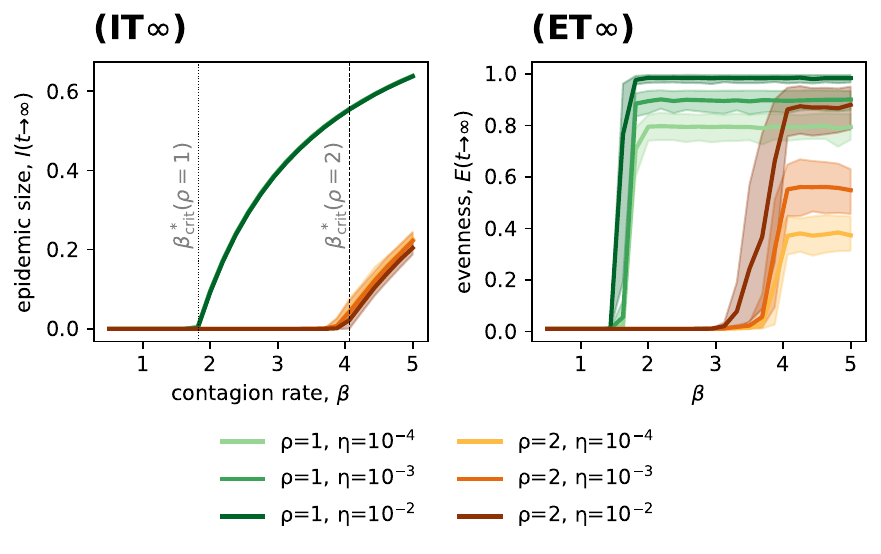} 
    \vspace{0.5cm}
    \end{minipage}
    \begin{minipage}{\textwidth}
      \includegraphics[width=\textwidth]{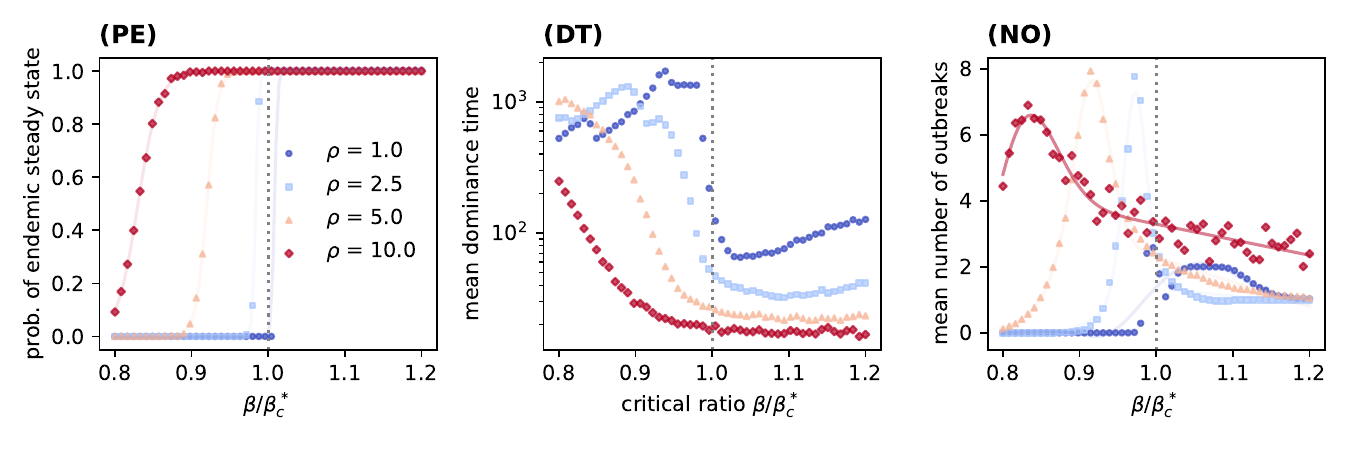} 
    \end{minipage}

    \caption{\textit{(Top)} Asymptotic epidemic size \(\textbf{(IT}\infty\)\textbf{)} and evenness \(\textbf{(ET}\infty\)\textbf{)} obtained varying \(\beta\) in a range comprising the estimated \(\beta_\text{crit}^*\), for \(\rho \in \{1,2\}\); envelopes represent 10\%-90\% quantiles. Results are obtained with \(M=100\) strains, and are averaged over 200 simulations, in which \(\mathbf Q\) (Erdos-Renyi, \(p_\text{conn} = 0.12\)) is resampled each time. \textit{(Bottom)} We analyze dynamics for the system \eqref{eq:model_homMF} in the critical region, \(\beta \approx \beta_\text{crit}^* \equiv \left\langle K\right\rangle^{-1}\), where \(\langle K \rangle \equiv M^{-1}\sum_{a,b}K_{ab}\), by computing three quantities of interest for different values of cross-immunity strength \(\rho\) by sampling 1000 realizations of the same system. \textbf{(PE)} The probability of reaching an endemic steady state at long times  quickly saturates to one as \(\beta \to \beta_\text{crit}^*\). \textbf{(DT)} Dominance time of a strain is defined as the length of the time interval during which the strain has the largest relative prevalence. The mean dominance time (MDT) is an indicator of the (inverse) rate of turnover between strains. For all values of \(\rho\), the MDT decreases sharply as \(\beta\) approaches \( \beta_\text{crit}^*\) from below. \textbf{(NO)} The mean number of outbreaks peaks as \(\beta \to \beta_\text{crit}^*\), and decreases in the super-critical region \(\beta > \beta_\text{crit}^*\), as large infectivity quickly drives the system to a steady state. Lines, where present, are mere interpolations, intended to aid visualization. Additional details are reported in Appendix~\ref{app:computational_details}.} 
    \label{fig:2}
\end{figure*}

More generally, whether \edit{some novel strains can invade over time, and whether the pathogen can become endemic}, depends dynamically on how the susceptibles distribute across the strain indices: concretely, a mutant strain \(a\) appearing at time \(t\) can grow in prevalence only provided that \(\dot I_a(t) / I_a(t) \approx - 1+ \beta\sum_b K_{ba}S_b(t) > 0\). This \edit{makes it impossible} to define a \edit{sharp} epidemic thresholds for the appearance of \edit{any given mutant strain independently of the history of the system}. However, it is still possible to define a \textit{critical region} in the proximity of \(\beta_{\text{crit}}^* = 1/\left\langle K\right\rangle\), \edit{where \(\langle K \rangle \equiv M^{-1}\sum_{a,b}K_{ab}\). This corresponds to the assumption that\footnote{\edit{We note that the relative prevalences of strains tend to be similar if the mutation rate \(\eta\) is large. In particular, as the mutation term becomes dominant --and provided that the mutation network is fully connected-- then \(I_a \approx I/M, S_a \approx S/M \ \forall a\), and eq.~\eqref{eq:model_homMF} reduces to one-strain SIS dynamics.}} \(S_a(t) = 1/M \ \forall a\) and that columns of \(K\) are statistically equivalent -- in particular, that their entries have the same mean.} This quantity is more informative than the invasion threshold \(\beta_\text{crit}\), \edit{as it marks the boundary of different regimes in which the system can or cannot reach an endemic state}. In particular, we find that as \(\beta\) crosses \(\beta_{\text{crit}}^*\) from below, the long-time epidemic size, \(I(t \to \infty)\), attains on average a non-zero value which increases with \(\beta\) (Fig.~\ref{fig:2}\textbf{(IT\(\infty\))}). A similar conclusion holds for the evenness of such endemic states, with sharply increases to its maximum values across the epidemic threshold; as expected, to larger mutation rates \(\eta\) correspond greater endemic diversity \(E(t \to \infty)\) (Fig.~\ref{fig:2}\textbf{(ET\(\infty\))}). As for the threshold itself, \(\beta_{\text{crit}}^*\) increases with \(\rho\) approximately as \(\beta_{\text{crit}}^* \sim e^{\rho}\) (see Appendix~\ref{app:diffusion_distance}).

Beyond asymptotic states, \(\beta_\text{crit}^*\) marks a regime shift \edit{in dynamics, as described by the rate} of strain turnover, number of transient outbreaks, and probability of the pathogen to reach a long-lasting endemic state (Fig.~\ref{fig:2}). By focusing our analysis to the critical region, we find that dynamics are qualitatively similar for widely different cross-immunity strengths. The probability \edit{(with respect to different realizations of the random mutation matrix \(Q\))} of reaching a stable endemic state increases sharply as \(\beta\) crosses \(\beta_\text{crit}^*\) (see Fig.~\ref{fig:2}\textbf{(PE})), as does the mean strain turnover rate (its inverse, the mean dominance time, is reported in Fig.~\ref{fig:2}\textbf{(DT})). Instead, the mean number of outbreaks within a given time horizon peaks for \(\beta \approx \beta_\text{crit}^*\) (Fig.~\ref{fig:2}\textbf{(NO)}). For larger values of \(\rho\), the structural details of the mutation matrix become relevant, as shown by an increased transition width in endemic probability and strong fluctuations in the number of outbreaks. Moreover, relative fluctuations in the entries of \(\mathbf K\) increase with \(\rho\); as a consequence, the actual epidemic threshold tends to anticipate the theoretical estimate of \(\beta_\text{crit}^*\), owing to the presence of strains characterized by a much more favorable cross-infectivity as the average value \(\langle K \rangle\).

\emph{Low-dimensional and directed mutation structures.} \edit{The effective \edit{antigenic} space of strains has been found empirically to be low-dimensional for flu~\cite{smith2004mapping} and Covid-19~\cite{wilks2023mapping}. To test whether this insight is compatible with our framework}, we set \(Q\) to be an \textit{exponential random geometric network}, where off-diagonal pairs of nodes have an exponentially decreasing probability of being connected, resulting in a \textit{quasi-one-dimensional} strain space; the broadness of the exponential profile describes the tendency of mutations to `jump ahead' in the considered strain sequence. Explicitly, the probability of nodes \(i\) and \(j\) to be connected is \(p_0 \exp\left(-rM|i-j|\right)\): \(p_0\) is the basic connection probability of two nodes (i.e. nodes with \(|i-j| \ll (rM)^{-1}\) are connected with probability \(\approx p_0\)), while \(r\) is a (inverse) broadness parameter determining the shape of the network relative to the number of strains. In the limit of this broadness encompassing the whole mutation network, i.e. \(r \ll 1/M\), we return to the Erdos-Renyi case, without clear-cut geometric features. In the opposite limit \(r \gg 1/M\), instead, a perfectly one-dimensional strain space is obtained, matching models of one-dimensional (and continuous, for large \(M\)) strain labels. As long as a general mutation network can present local one-dimensional structures, this hypothesis is reasonable at appropriate scales, and it makes said patterns of sharp strain turnover more likely to be observed.

In general, however, with this configuration, the system often reaches an endemic state characterized by high strain diversity. As the existence of such states is not supported by empirical evidence, these should rather be considered as descriptive of the long-term outcome of pathogen evolution, which is analyzed later in the paper. 

To better match the expected empirical patterns \edit{for pathogens of interest, such as the ones for flu or COVID-19}, we enforce the exclusion of earlier strains, which is motivated by an effective accounting of multiple immune memories as detailed in Appendix~\ref{app:directed_mutations}. Specifically, we impose that \(\mathbf Q\) and \(\mathbf K\) are triangular matrices, reflecting a temporal ordering of strain emergence: this ensures that the fitness of older strains decreases with time, and that these cannot reappear by back-mutation of newer strains, thus preventing their resurgence.

The practical relevance of the previous argument becomes evident in comparing the model predictions to empirical data on the circulation of COVID-19 strains~\cite{hodcroft2021covariants}. We assume \(\beta = \beta_\text{crit}^*\). This is justified through the biological insight that selective pressures drive infectivity towards the rate which gives better chances of long-term survival of a given strain~\cite{kao2006evolution,lion2018beyond}. As people infected with COVID-19 are highly unlikely to be infective after 10 days from the onset of symptoms~\cite{walsh2020duration}, we assume for the typical recovery time (i.e. the implicit time unit of our framework) a conservative estimate of one week, which simplifies analysis as it aligns with weakly frequency of prevalence estimates. By a straightforward scan over the parameter space, we identify configurations which align to empirical data in terms of mean dominance time and peak epidemic size. 

\begin{figure*}
    \centering
    \includegraphics[width=\textwidth]{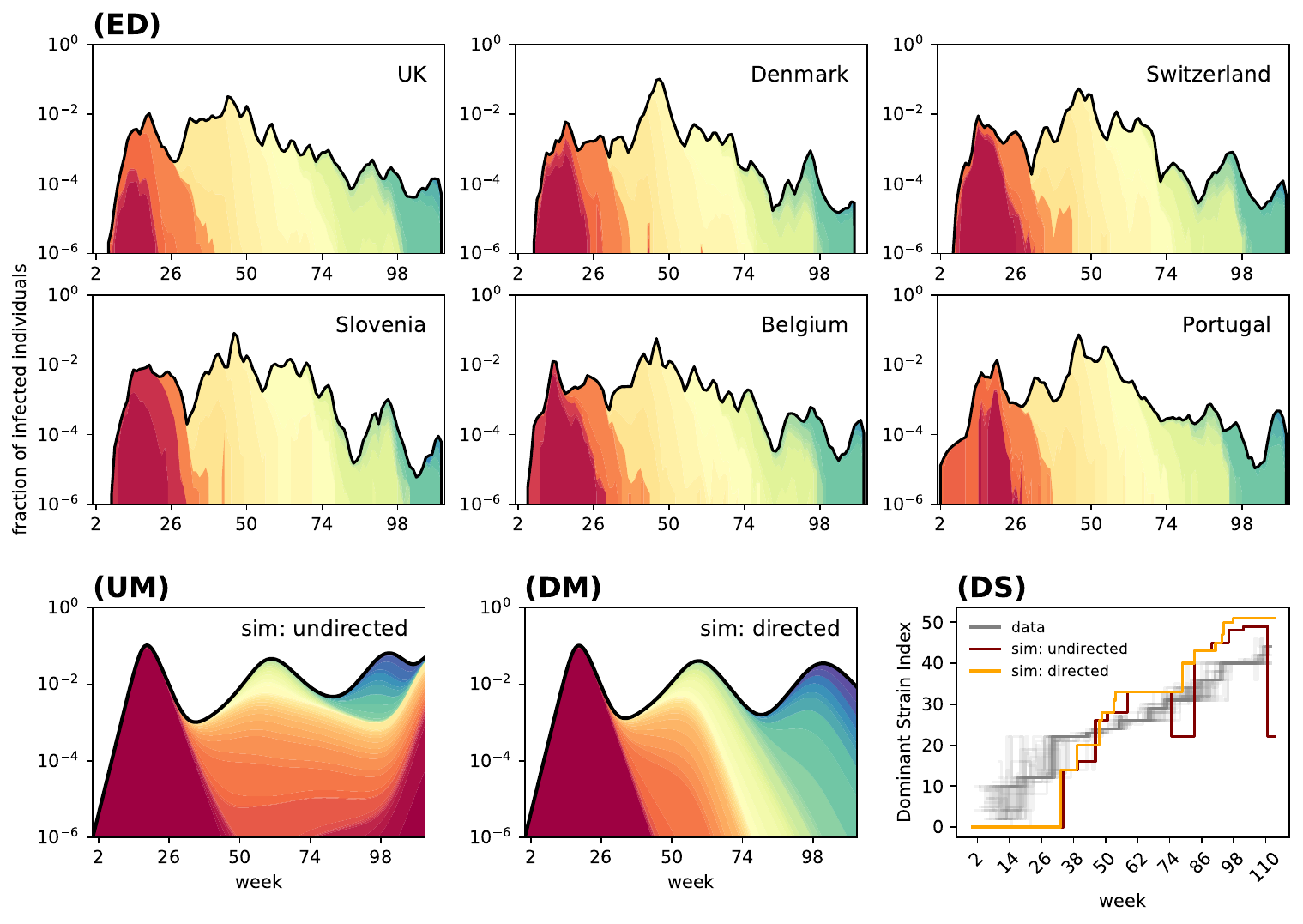} 
    \caption{We compare evo-SIS model~\eqref{eq:model_homMF} predictions \textbf{(UM, DM, DS)} with empirical data on COVID-19~\cite{hodcroft2021covariants} \textbf{(ED)}, where \(M=52\) strains are observed; six countries are shown as examples. \edit{In all plots, different colors denote the relative prevalence of different strains}. We identify likely model parameters by matching mean dominance time and peak epidemic size predicted by the model to the empirical values, calculated globally, which measure respectively \(5.42\pm1.69\) weeks and \(\left(3.99\pm3.07\right)\times 10^{-2}\) (relative to total population). We report dynamics with a configuration matching these empirical values, both with a directed \textbf{(DM)} and undirected \textbf{(UM)} mutation structure. We also compare the dominant strain trajectory \textbf{(DS)}: the undirected case shows resurgence of earlier strains which are not observed in data, while the directed cases --after clearing the earlier strain-- shows a more realistic progression. Additional details are reported in Appendix~\ref{app:computational_details}.} 
    \label{fig:4_data_sim}
\end{figure*}

In Fig.~\ref{fig:4_data_sim} we compare model prediction and time series for the COVID-19 epidemic observed in several countries. We note that the estimated number of infected individuals vary over at least four orders of magnitude. \edit{Across different countries, we observe a robust patterns of a few (two to three) prominent peaks in epidemic size, with secondary likely deriving from events which we regard as random perturbation of a mean behavior.} Our model incorporates no mitigative effects such as social restrictions or vaccinations, nor does it take into account finite size effects, but it is able to reproduce the broad characteristic of empirical patterns, with successive peaks and a steady turnover of dominant strains. In this regard, we notice that the earliest strain, in our model, remains dominant for a much longer time than observed. However, we must remember the aforementioned limitations of the model. The well-mixedness assumption is also a very broad approximation for country-level data, especially so in scenarios of limited contact rates such as during lockdowns. It appears likely that these effects might have contributed to moderating oubreaks' sizes\footnote{We also note that the data we used have been cleared of unidentified or recombinant strains.}. In this sense, the predictions of our simple models represent a null scenario with respect to the realized epidemic dynamics of mutating pathogens, excluding the necessity of further mechanisms to explain the coarse characteristics of empirical observations.

\vspace{0.5cm}

\subsection{Near steady-state within-deme evolution}  \edit{We now turn to the analysis of dynamics near steady states. This is an appropriate description for endemic pathogens, whose prevalence in the population is approximately constant. Moreover, within the domain of validity of this approximation, the following construction elucidates how selective pressures shape diversity patterns.}

Following Appendix~\ref{app:time_scale_separation}, dynamics in eq.~\eqref{eq:model_full} or ~\eqref{eq:model_homMF}  can be recast as a set of integro-differential equations by integrating out the susceptible compartments altogether. Further, we assume that FoIs are large: \edit{in the single-deme case, these are defined as \(\lambda_{a}(t) = \beta\sum_b K_{ab}I_b(t) = \beta I(t) K_{ab}p_b(t)\), where \(I(t) = \sum_aI_a(t)\) is the epidemic size at time \(t\)}. In this regime, we can perform an an adiabatic series expansion~\cite{lugiato1984adiabatic} (ASE) of dynamics up to any desired order in powers of \(\lambda_a (t)\). \edit{Truncating the ASE at first order coincides with the more commonplace quasi-stationary approximation, which holds near steady states, while higher-order terms can capture transient dynamics.}

At leading-order (LO) of the ASE, susceptibles are completely constrained to their quasi-stationary value: \(S_a = I_a / \lambda_a\); after plugging this into eq.~\eqref{eq:model_full1}, dynamics is expressed in terms of relative prevalence as

\begin{equation}\label{eq:model_homogeneous_RM}
    \dot{p}_{a} = \phi^a p_{a} - \eta \left(\mathbf L_Q p\right)_{a} \ ,
\end{equation}

where the last term is a shorthand for matrix-vector multiplication. \edit{This equation holds up to order \(\left[\inf_{a,t} \lambda_a(t)\right]^{-1}\).} Here, we introduce the fitness factor,

\begin{equation}\label{eq:fitness_tensor_homogeneous}
    \phi^{\,a}(p):=\sum_b \frac{K_{b a}\,p_b}{(\mathbf K p)_b}-1 \ .
\end{equation}

Thus the fitness factor, which encodes selective forces ensuing from inter-strain competition, depends on the full strain composition. Eq.~\eqref{eq:model_homogeneous_RM} is a replicator-mutator-like (RM) dynamics~\cite{komarova2004replicator}, where the fitness factor provides a strain-specific correction to the diagonal of the Laplacian matrix \(L_Q\), thus modifying the pure random walk mutation process.

The LO dynamics predicts that epidemic size does not change over time. To obtain an explicit feedback between epidemic size and strain composition, we calculate the next-to-leading-order (NLO) terms in the ASE, finding

\begin{subequations}\label{eq:model_homogeneous_2nd_order}
 \begin{multline}\label{eq:model_homogeneous_2nd_order1}
\dot p_a =
p_a\phi^{a}(p) - \eta(\mathbf L_Q p)_a
\\+ \frac{p_a}{\beta I}\sum_b\left[\frac{(\mathbf K p)_b - K_{b a}}{(\mathbf K p)_b^{2}}\left(p_b\phi^{b}(p)-\eta(\mathbf L_Q p)_b\right)\right]
\\+\mathcal{O}(\beta^{-2}) \ ,
\end{multline} 
 \begin{equation}\label{eq:model_homogeneous_2nd_order2}
    \dot{I}
=
-\sum_b \frac{1}{\beta (\mathbf K p)_b} \left(p_b\phi^{b}(p) - \eta(\mathbf L_Q p)_b\right)+\mathcal{O}(\beta^{-2}) \ .
\end{equation}
  
\end{subequations}

This reformulation of the evo-SIS dynamics brings to light the two following facts. 

First, in the dynamics of relative prevalences, eq.~\eqref{eq:model_homogeneous_2nd_order1}, NLO terms become increasingly less important as the epidemic size \(I\) increases: if the epidemic size reaches a stable large value, then the LO dynamics, eq.~\eqref{eq:model_homogeneous_RM}, drives the system near the steady state. Hence, the system evolves under both \edit{mutation drift} and selective forces, which however act additively. Conversely, NLO terms become large when \(I\) is small: we note that, in the r.h.s. of eq.~\eqref{eq:model_homogeneous_2nd_order1}, the effective terms couple multiplicatively selection and mutations. \edit{We note that these terms describe an explicit form of antigenic competition which is readily interpretable. Inside the square brackets in eq.~\eqref{eq:model_homogeneous_2nd_order1}, the factor \(\left(\left(\mathbf K p\right)_b - K_{ba}\right)\left(\mathbf K p\right)_b^{-2}\) compares the cross-infectivity of strains \(a\) and \(b\) with the instantaneous average of the cross-infectivity of \(b\) with all strains. This factor multiplies the LO dynamics of \(p_b\), hence whether the growth in relative prevalence of each strains enhances or impairs the growth of strain \(a\) depends on the cross-infectivity of the full set of strains. Moreover, by collecting \(p_a\) in eq.~\eqref{eq:model_homogeneous_2nd_order1}, the factor \((\beta I)^{-1}\sum_b[\dots]\) can be seen as a NLO contribution to the fitness, which thus accounts explicitly for the effects of mutations on selection, beyond mere diffusion in strain space.}

Second, at NLO, epidemic size is entirely driven by the strain composition. In fact, the two terms appearing in the r.h.s. of eq.~\eqref{eq:model_homogeneous_2nd_order2}, which are weighted against each strain susceptibility, are similar to those appearing in eq.~\eqref{eq:model_homogeneous_2nd_order1} and subject to the same interpretation. We note that the r.h.s. of eq.~\eqref{eq:model_homogeneous_2nd_order2} tends to be positive if strains which are decreasing in prevalence (\(\dot p_b < 0\) at LO) \edit{do so while the corresponding susceptible individuals are no longer available for new infections, (\(\left(\mathbf  K p\right)_b \to 0\)), signaling that the epidemic dynamics is shifting towards different strains. In other words, epidemic growth is primarily driven by shifts in the antigenic profile of the host population.}

As shown in Fig.~\ref{fig:4}\textbf{(E,TI)}, successive approximant dynamics are in excellent agreement with the full system~\eqref{eq:model_full}, provided that initial conditions belong to an appropriate manifold. Hence, these dynamics are accurate in a non-trivial neighborhood of the steady state. NLO and NNLO corrections bring marginal improvements in accuracy. LO description provides a good compromise between simplicity and accuracy at the level of relative prevalence near the steady state, while NLO extends the LO insights to epidemic transients. We note that the quasi-stationary regime, while streamlining dynamics, clearly retains the dynamical patterns observed in invasion dynamics, such as a peak in evenness observed as the epidemic size nears its maximum.

\subsection{Evolutionary landscape in a metapopulation network} 

\edit{In this section, we consider a more complex model, moving from the single-deme discussed before to the whole metapopulation network, introduced in \eqref{eq:model_full}.}

\edit{As anticipated earlier, the metapopulation network allows us to describe different scenarios, from geographic patches with a structure encoding human movements to more structured populations stratified by available features (age, socio-economic class, etc.)~\citep{gomez2018critical,soriano2018spreading,arenas2020modeling,mistry2021inferring,manna2024generalized,maniscalco2025critical}. Furthermore,} by considering an heterogeneous metapopulation structure, we can reveal how spatial and mutational effects contribute to the evolutionary landscape, \edit{which is one of the main contributions of our study}. 

The ASE proceeds as in the previous paragraph, and Appendix~\ref{app:time_scale_separation}. The general structure of LO dynamics is the same as in the homogeneous case\footnote{As in the homogeneous case, the epidemic size \(I(t) = \sum_{x,a}I_{xa}(t)\) does not change in time at LO, with dynamics being recovered at NLO and beyond.}: writing relative prevalence at node \(x\) for strain \(a\) as \(p = \left\{p_{xa}\right\}_{x,a}\), where \(p_{xa} = I_{xa}/\sum_{x,a} I_{xa}\), we have

\begin{equation}\label{eq:model_RM}
    \dot{p}_{xa} = \sum_y \phi_{xy}^a p_{ya} - \eta \left(\mathbf L_Q p\right)_{xa} \ ,
\end{equation}

where the last term is a shorthand for matrix-vector multiplication in which the contact index plays no role. Strain fitness is now encoded in the tensor

\begin{subequations}\label{eq:fitness_tensor}
    \begin{equation}\label{eq:fitness_tensor1}
        \phi_{xy}^a(p) = A_{xy}R_{xa} - \delta_{xy} 
    \end{equation}
    \begin{equation}\label{eq:fitness_tensor2}
        R_{xa} = \sum_b \frac{K_{ba}p_{xb}}{\sum_{z,c} A_{xz}K_{bc}p_{zc}} \ .
    \end{equation}  
\end{subequations}

The fitness tensor comprises a strain- and space-dependent factor \(R_{xa}\), which ties together cross-immunity and spatial effects, as different strains directly compete through adjacent \edit{demes}. \edit{More specifically, \(R_{xa}\) weights the flow of relative prevalences, in the neighborhood of deme \(x\), from each strain \(b \neq a\) into strain \(a\) against the flow from \(b\) to any other strain. Hence inter-strain competition now takes into account the fact that the full host population is not well-mixed, making the description dependent on the metapopulation structure.}

\edit{Interstingly, eq.~\eqref{eq:model_RM} has the form of a modified random walk that depends on the interplay between the metapopulation network and the mutation network}. In fact, we can recast eq.~\eqref{eq:model_RM} in terms of an effective \textit{dynamical evolutionary landscape} \(\mathbf \Gamma\) as 

\begin{equation}\label{eq:multiplex_dynamics}
    \dot p_{xa} = \sum_{yb}\Gamma_{xa}^{yb}p_{yb} \ , \quad \Gamma_{xa}^{yb} = \left(\phi_{xy}^a \delta_{ab} - \eta {L_Q}_{ab}\delta_{xy} \right) \ .
\end{equation}

\edit{Note that, formally, \(\mathbf \Gamma\) acts as a (state-dependent, hence adaptive) Laplacian over the multilevel structure of mutations and metapopulations. In this case}, the fitness tensor and mutation Laplacian which enter \(\mathbf \Gamma\) do not mix the two different spaces\edit{, as the former is diagonal in the strain space, and the latter is diagonal in the metapopulation structure}. In regimes where different strains all have near-zero fitness, \(\mathbf \Gamma\) reduces to the mutation Laplacian, local and identical in each \edit{deme}, while in the limit \(\eta \to 0\), prevalence dynamics is \edit{driven by antigenic selection acting on a set of already present strains, with novel strains appearing rarely}. We remark that the above formulation of the landscape is well-defined in the limit where adiabatic elimination of susceptibles holds at leading order. In departing from this regime, one should include higher order terms, or revert to utilizing a fully dynamical representation of \(S_{xa}\). Hence, we will limit our analysis the landscape connectivity near steady states.

The preceding analysis assumes \edit{demes} are identical, aside from their location within the \edit{metapopulation} structure. However, an empirical host population is itself heterogeneous, and could be partitioned in subgroups according to the interaction of individual hosts with the pathogen.  Hence, we consider \edit{deme}-specific mutation matrices, \(\mathbf Q \to \mathbf Q^x\); as a consequence, \(\mathbf K \to \mathbf K^x \) is also \edit{deme}-specific. The interplay of the metapopulation-mutation structures in \(\mathbf \Gamma\) has the important consequence of creating bridges between different connected components in the mutation structure: even if a given mutation path is interrupted in a given \edit{deme, it may be resumed upon spreading in a different location}. Hence, a \edit{well interconnected} metapopulation structure can create the conditions for pathogens to more freely explore the entire mutational space, bypassing \edit{deme-specific immune defenses}.

\begin{figure}

    \centering
    \includegraphics[width=1\columnwidth]{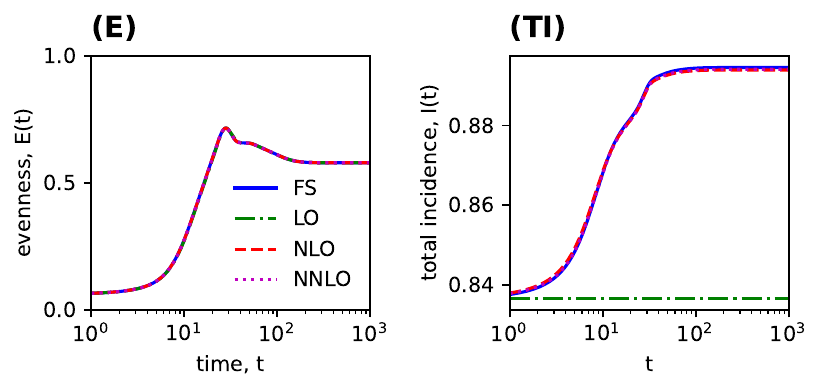}
    \includegraphics[width=1\columnwidth]{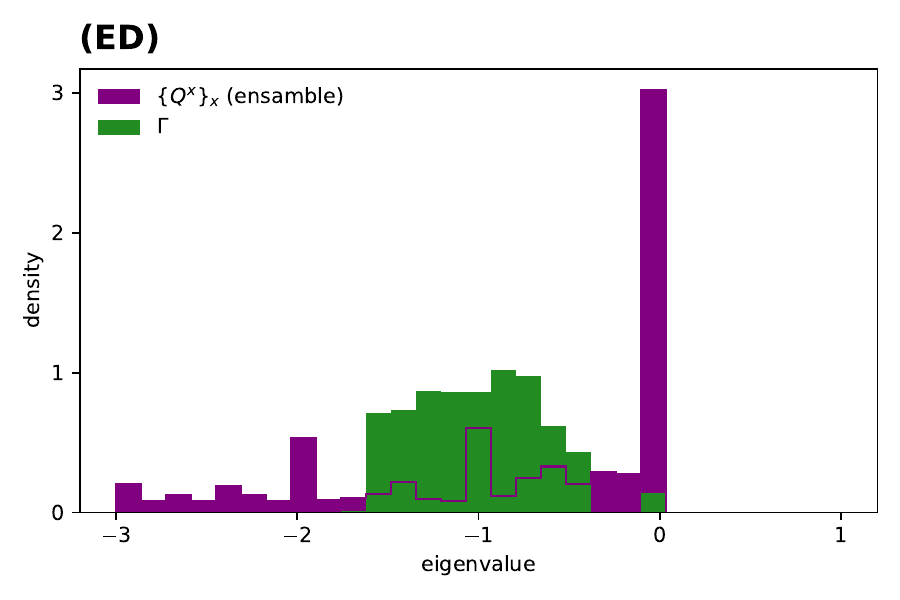}
    \includegraphics[width=1\columnwidth]{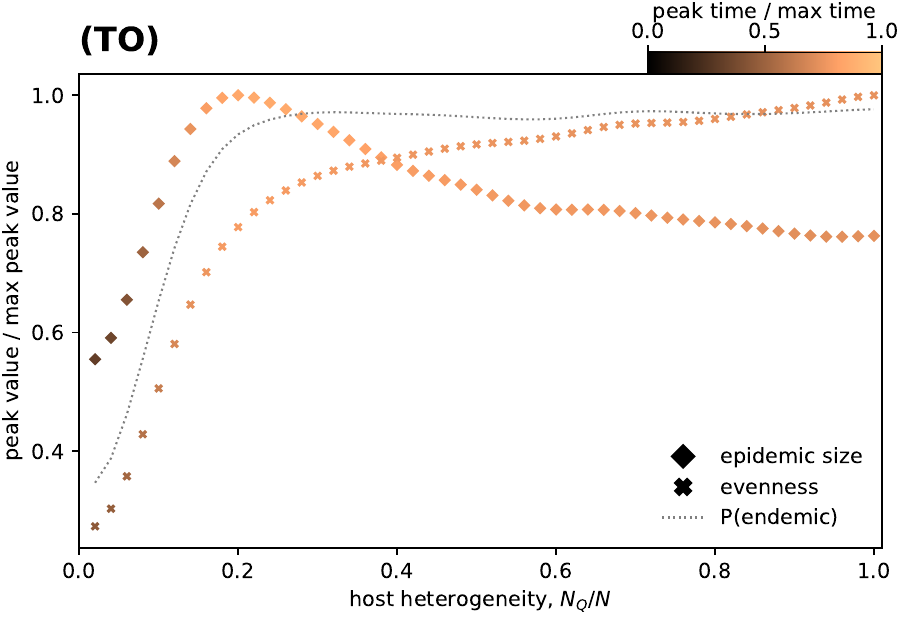}  
    
    \caption{Within a single well-mixed deme, in a regime of large transmission rate and weak cross-immunity (\(\beta = 21,\rho = 1.4\)), the complete dynamics and its approximants give near identical descriptions of relative abundances, and thus of evenness \textbf{(E)}. Leading order dynamics fail to predict variations in total incidence \textbf{(TI)}, which is however recovered by higher-order approximants. Instead, for a population of heterogeneous demes, we compare eigenvalue density \textbf{(ED)} of the steady state evolutionary landscape \eqref{eq:multiplex_dynamics},  \(\left.\mathbf \Gamma\right|_{t \to +\infty}\), to that of the ensamble of local mutation matrices \(\mathbf Q^x\): a spectral gap emerge in the former which is lacking in the latter, indicating the emergence of an interconnected structure. \textbf{(TO)} Strains' evenness increases monotonically with host heterogeneity \(N_Q/N\), while epidemic size reaches maximal values for an non-trivial, intermediate value. The plot reports values normalized by their maximum attained value over the full range of \(N_Q/N\), in order to display variations on comparable scales. The metapopulation structure is modeled with an Erdos-Renyi network with \(p_\text{conn}^A=0.15\), \(N =50\), and one connected component. Additional details are reported in Appendix~\ref{app:computational_details}.}
    \label{fig:4}
\end{figure}

\edit{To demonstrate the previous assertions,} we analyze numerically this regime by again assuming an exponential random geometric ensemble of mutation matrices. This creates an approximately one-dimensional space (in the sense discussed in previous sections) for \edit{strains to explore}, without the extra structure encoded in e.g. hierarchical block models, allowing us to disentangle effects of coarse topology from structural properties such as nestedness. As illustrated in Fig.~\ref{fig:4}\textbf{(ED,TO)}, it is apparent how a mutation network fragmented in several components can be effectively reconnected by spreading between different demes, where diversification leads to successful antigenic escape: a pathogen struggling to become endemic in a homogeneous deme instead achieves large incidence over time in a more heterogeneous population, with similarly large strain diversity. Indeed, a spectral comparison between \(\mathbf \Gamma\) and the ensemble sample \(\{\mathbf Q^x\}_x\) reveals how the former, in stationary regime, provide robust connectivity --as \edit{suggested} by a gap between bulk and near-zero spectrum-- which is absent in the ensemble (Fig.~\ref{fig:4}\textbf{(ED)}). Moreover, we simulate different degrees of heterogeneity by sampling the local mutation matrices from pools of size \(N_Q\), so that the ratio \(N_Q / N \in (0,1]\) quantifies host heterogeneity\edit{; note that \(N_Q = 1\) correspond to the case of identical demes}. Our results confirm the intuitive idea (Fig.~\ref{fig:4}\textbf{(TO)}) that increased host heterogeneity boosts strain diversity by enabling extended evolutionary paths. Average epidemic size, instead, is maximized by a value of \(N_Q / N\) in the interior of the range \((0,1]\). This is to be compared with the probability of reaching an endemic state increases with heterogeneity from small values, and saturates at about 0.9 for \(N_Q / N \gtrapprox 0.2\). \edit{As a broader mutation landscape correlates with an immune response which varies similarly across different demes, increased strain variety balances off against greater obstacles in spreading.} These results indicate that host heterogeneity enables a trade-off between epidemic size and pathogen diversity over long time scales.

\section*{Discussion and conclusions}

In epidemic modeling, \edit{the coevolution of pathogens and host immunity  is usually studied in detail for only one of the two processes, while the other neglected or greatly simplified.}
In particular, pathogen mutations are commonly assumed to explore a homogeneous --and in fact, continuous-- strain space. \edit{However, as different strains are identified by their genetic sequences, one should more generally adopt a likewise categorical description of their identities. Moreover, while the antigenic space of strains reconstructed in the literature is found to be low-dimensional, it is not clear whether this results holds \emph{a priori}, instead of emerging from an interplay of different processes (e.g. epidemiological, evolutionary, or related to the sampling of strains). To investigate the possible origins of the low-dimensionality of the strain space, a framework in which this property is not assumed a priori should be utilized. \emph{A fortiori}, given that we are considering a categorical space, a global notion of dimensionality could generally be ill-defined. Crucially, since the antigenic similarity of different strains should be measured by taking into accounts the properties of the strain space, the immune response of hosts --and, in particular, cross-immunity phenomena-- should also reflect these considerations.} 

In this work, \edit{complementarily to previous studies}, we considered an evo-SIS model in which different pathogen strains are nodes of a network, linked by mutations which change the type of dominant strain. \edit{The key element of this framework is description of cross-immunity effects, which enables host-mediated competitive interactions between different strains.} We assumed hosts possess perfect immune protection from the latest strain they encountered, and partial immunity from other strains according to antigenic similarity; we described strains which are connected by many short mutational paths as similar.

In a first step, we assumed that the host population is well-mixed. Already at this level, invasion phenomenology is qualitatively compatible with empirical epidemiological patterns, with consecutive outbreaks driven by inter-strain competition mediated by population-wide antigenic response. In between outbreaks, strain composition adapts dynamically to the immune profile of hosts', which is determined by the previous epidemics. The pathogen evolves to probe the hosts' population defenses: if at least one strain can penetrate the antigenic barrier, a new outbreak will take place. This phenomenology integrates boom-and-bust dynamics typical of SIR models with the strain diversification of replicator-mutator dynamics. \edit{We show the importance of the topological properties of the mutation network, which is necessary to make contact with empirical data concerning epidemics of quickly-mutating pathogens. To validate our theoretical analysis, we considered the COVID-19 epidemic. Under the assumption of well-mixed population, we compare multi-strain data~\cite{hodcroft2021covariants} obtained in different countries to evo-SIS predictions, both for undirected and directed structures. We find that the model, despite its simplicity (and lack of external inputs, such as vaccinations and lockdowns) recovers the general features of the COVID-19 epidemic. However, this is the case only for a directed mutation structure. Nevertheless, the undirected framework remains relevant for pathogens in which strain resurgence is actually observed.}

In the long time limit the system settles in an evolutionary and epidemiological steady state, where the pathogen either goes extinct or becomes endemic. In the latter case, within this framework of homogeneous host population, we find that the long time strain diversity is often smaller than the peak diversity, which is achieved transiently during an epidemic phase. Near endemic steady states, our model is amenable to a perturbative scheme which brings out a structure similar to that of replicator-mutator dynamics, and couples changes in strain diversity and epidemic size. Higher-order contributions from this scheme bring better accuracy, but the first two orders are sufficient to identify the effective processes responsible for some striking features observed in numerical experiments. Importantly, these effective terms clearly encapsulate selection driven by strain competition and antigenic escape.

In its complete form, our framework recognizes the heterogeneity of hosts' population, both in terms of metapopulation structure and in their physiological differences. We can account for differences between local populations through a stochastic approach: to each deme corresponds a different realization of the same mutation structure. Heterogeneous hosts have been considered before in minimalist models, e.g. to account for immunodeficient individuals~\cite{kumata2022antigenic}; with a broader scope, this is typically done by employing randomly distributed parameters~\cite{novozhilov2012epidemiological}. Our work generalizes this approach, and adds new insight by demonstrating the emergence of a dynamical evolutionary landscape, whereupon pathogen spreading through different demes can open new evolutionary pathways. As the emergence of new strains with increased lethality poses a serious risk to public health~\cite{morens2008emerging}, this insight suggests the paramount importance of epidemic evolutionary containment by means of structural curtailing of metapopulation networks during periods of high pathogen circulation. 

In a future planned work, building on the argument presented in Appendix~\ref{app:RM_within_host}, we will model in greater detail the mutation dynamics of pathogens within hosts, where immunity can be better described in mechanistic terms. This will constrain our choice of mapping between mutational structure and antigenic similarity; additionally, instead of treating each individual host as infected with one strain at the time, it will appropriately take into account lineage divergence and co-infection in a comprehensive framework. \edit{Further, we will leverage the heterogeneous description of immune coverage built into our evo-SIS framework to predict the outcome of vaccination strategies in light of cross-strain interactions~\cite{jiang2026learning}.}

\section*{Acknowledgments}

\edit{DZ and VB acknowledge financial support from the Human Frontier Science Program Organization (HFSP Ref. RGY0064/2022). SA and MDD acknowledge financial support from the project INFN-LINCOLN. MDD acknowledges partial financial support from from Human Frontier Science Program Organization (HFSP Ref. RGY0064/2022), from the EU’s Horizon Europe research and innovation programme under grant agreement No. 101186013 (VirHox), and from the MUR - PNC (DD n. 1511 30-09-2022) Project no. PNC0000002, DigitAl lifelong pRevEntion (DARE). The authors thank Valeria D'Andrea and Charley Presigny for their useful comments.}

\appendix

\section{Within-host replicator-mutator dynamics for a population of pathogen strains}\label{app:RM_within_host}

As noted in the main text, our choice in modelling pathogen mutation is is different from other approaches found in the literature. In our framework, instead of using replicator-mutator dynamics, with mutations taking place during infection events, we describe mutations as a random walk (RW) of the infected host's strain label. This corresponds to assuming that the dominant strain within each host undergoes itself a random walk. In this Appendix, we illustrate through an auxiliary model the hypotheses leading to our modeling choice. Moreover, we connect the \textit{macroscopic} (i.e. host-level) mutation matrix \(\mathbf Q\) to its \textit{microscopic} (i.e. pathogen-level) counterpart, which we denote \(\mathbf m\) in the following.

Writing \(x_a\) for the proportion of strain \(a\), we consider a replicator-mutator (RM) dynamics in which the fitness of each strain decays according to its own recent prevalence, representing the effect of host immune response. Denoting by \(x_a(t)\) the frequency of strain \(a\) among all \(M\) circulating strains at time \(t\), the dynamics written in dimensionless units reads
\begin{equation}\label{eq:RM_WH_dynamics}
    \dot{x}_a = \sum_{b} f_b(x,t)\,x_b\,m_{ab} - x_a\,\bar{f}, 
    \qquad 
    \bar{f} = \sum_{c} f_c(x,t)\,x_c ,
\end{equation}
where \(f_a(x,t)\) is the instantaneous fitness of strain \(a\). We introduce a \emph{memory–dependent} fitness,
\begin{subequations}\label{eq:RM_WH_fitness_memory}
\begin{equation}
 f_a(x,t) = x_a(t)\exp\!\big[-\kappa\,\bar{x}_a^{(\tau)}(t)\big],   
\end{equation}
\begin{equation}
 \bar{x}_a^{(\tau)}(t)
      = \frac{1}{\tau}\!\!\int_{\max(0,t-\tau)}^{t}\!\!x_a(s)\,ds,    
\end{equation}
\end{subequations}
where \(\kappa>0\) sets the strength of immune suppression and \(\tau>0\) the time scale upon which immune memory produces an antigenic response\footnote{Note that \(\lim_{\tau \to 0^+}\bar{x}_a^{(\tau)}(t) = \kappa x_a(t)\). The opposite limit of \(\tau \to \infty\) gives \(\bar{x}_a^{(\tau)}(t) \equiv 0\) for any finite \(t\) unless also \(\kappa \to \infty\), with fixed ratio \(\kappa/\tau \equiv \delta\), so that immune memory is infinitely long: this leads to the permanent exclusion of older strains.  This can serve as a model for the situations which we consider in Appendix~\ref{app:directed_mutations}.}.  Each strain’s reproduction rate is therefore proportional to its current abundance \(x_a\) and exponentially penalized by its recent time–averaged prevalence \(\bar{x}_a^{(\tau)}\). This produces delayed self–inhibition: a strain that has dominated the population for a time \(\sim\tau\) experiences a vanishing fitness \(f_a\simeq 0\), opening the way for invasion by mutants. The structure of the mutation matrix \(\mathbf m\) determines how such invasions take place. \(\mathbf m\) describes reproduction followed by mutation, with elements \(m_{ab}\ge 0\) and \(\sum_a m_{ab}=1\). Diagonal elements \(m_{aa}=\mathcal{O}(1)\) represent faithful replication, while off–diagonal elements \(m_{ab}=\mathcal{O}(\varepsilon)\ll1\) encode small probabilities of mutation per replication.

Suppose strain \(a\) is currently the dominant one, so \(x_a\simeq 1\) and \(x_{b\neq a}\ll 1\). The mutation term then continuously produces small amounts of mutant units at rates \(\propto m_{ba}\). We assume that both \(\tau\) and \(\kappa\) are large, so that (i) \(a\) is initially unaffected by immune response, and (ii) the fitness of \(a\) becomes essentially null after memory kicks in. By these assumptions, we can consider three dynamical phases of strain turnover.

\vspace{0.3cm}
\noindent\emph{Phase I – establishment of the dominant strain.} 
When a strain \(a\) becomes dominant, e.g. after previous mutations or following host infection, \(\bar{x}_a^{(\tau)}\ll 1\), hence \(f_a \simeq x_a \simeq  1\) and \(\bar f\simeq f_a x_a \simeq 1\). Mutants satisfy
\begin{equation}
    \dot{x}_b \simeq m_{ba}f_a x_a - x_b \bar f 
                 \simeq m_{ba}-x_b.
\end{equation}
Thus each mutant \(b\) quickly relaxes to a small stationary abundance \(\sim m_{ba}=\mathcal{O}(\varepsilon)\). The total mutant mass \(\sum_{b\neq a}x_b\) therefore remains \(\mathcal{O}((M-1)\varepsilon)\).

\vspace{0.3cm}
\noindent\emph{Phase II – decay of the dominant strain fitness.}
As \(x_a(t)\simeq1\) persists over a memory window \(\tau\), the average \(\bar{x}_a^{(\tau)}\) increases and \(f_a(t)\) decreases exponentially, \(f_a\simeq e^{-\kappa \min\left\{t/\tau,1\right\}}\). Consequently, the effective reproduction term \(f_a x_a\) decays as \(x_a^2 e^{-\kappa \min\left\{t/\tau,1\right\}}\), causing a slow transfer of mass away from the dominant strain. Mutants remain at quasi–steady abundances until the fitness of \(a\) becomes negligible. At the end of this phase, \(f_a = e^{-\kappa}\approx 0\), and the expected residence time of each strain is \(\mathcal{O}(\tau/\kappa)\).

\vspace{0.3cm}
\noindent\emph{Phase III – invasion and replacement.}
When \(f_a\simeq 0\), seeding from \(a\) stops and all \(x_{b\neq a}\) evolve approximately according to
\begin{equation}
    \dot{x}_b \simeq x_b^2m_{bb}e^{-\kappa \bar{x}_b^{(\tau)}} - x_b\sum_c x_c^2 e^{-\kappa \bar{x}_c^{(\tau)}} .
\end{equation}
Among these mutants, the ones with \(\bar{x}_b^{(\tau)}\approx 0\) retain unsuppressed fitness and grow fastest, but we should expect this condition to hold rather uniformly for all mutants if their growth was uniformly suppressed during Phase I and if Phase II unfolded over a short time interval. Instead, the strain \(b^*\) with the largest product \(m_{b^*a}m_{b^* b^*}\) will typically reach macroscopic abundance first and become the new dominant strain, by the combined effects of low mutation rate and the initial advantage given by the frequent mutation from strain \(a\) during Phase II. Hence, the transition time between successive dominant strains scales inversely with \(m_{b^*a} m_{b^* b^*}\). After each replacement, the same sequence of dynamical phases repeats anew. These simple observations motivates the following analysis.

\begin{figure}
    \centering
    \includegraphics[width=1\columnwidth]{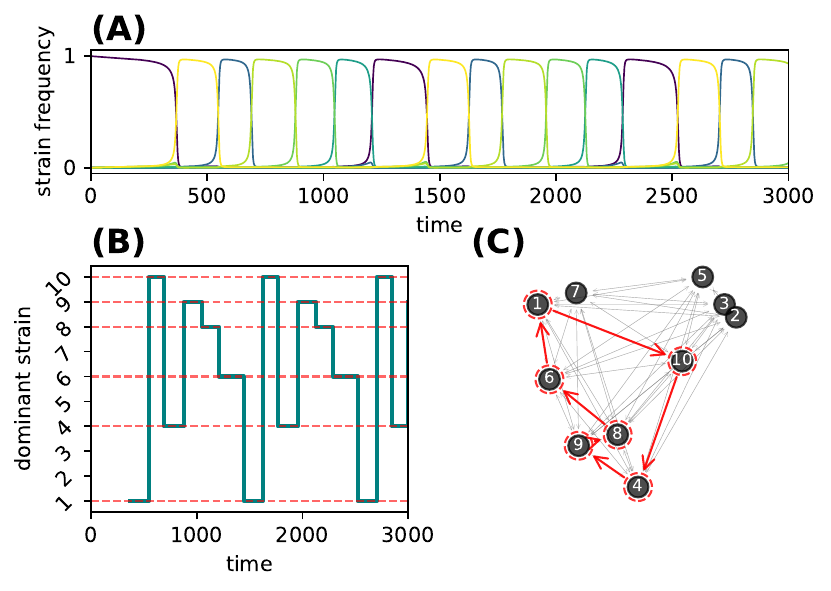}
    \caption{Deterministic cycles of within-host replicator-mutator dynamics, eq.~\eqref{eq:RM_WH_dynamics}, with an undirected mutation matrix \(\mathbf m\). We consider 10 strains, a mutation rate \(\varepsilon = 3\times 10^{-3}\), and off-diagonal entries of \(\mathbf m\) as equal to \(\varepsilon\), multiplied by a random number drawn by an exponential distribution of mean equal to 3. \textbf{(A)} Strain frequencies follow a recurrent pattern, with dominant strains reappearing in cycles \textbf{(B)}. At this deterministic level, these are identified \textbf{(C)} by the deterministic mapping in eq.~\eqref{eq:within_host_deterministic_successor}. Dashed lines identify predicted dominant strains both in \textbf{(B)} and \textbf{(C)}.}
    \label{fig:5}
\end{figure}

\vspace{0.6cm}
\noindent\emph{Deterministic cycling and macroscopic transition network.}
For fixed parameters \((\kappa,\tau,\varepsilon)\), given a mutation matrix \(M\), the replacement rule 

\begin{equation}\label{eq:within_host_deterministic_successor}
a\mapsto s(a) = b^* \in \arg\max_{b\ne a}(m_{ba}m_{bb})
\end{equation}

defines a deterministic map \(s(a)\) on the discrete set of strains. In the general case of \(\mathbf m\) being non-symmetric, outside of degenerate cases\footnote{Cases in which successor strains are not uniquely determined by the map in eq.~\eqref{eq:within_host_deterministic_successor} do in fact happen, less frequently so if off-diagonal entries of the mutation matrix are more heterogeneous. Interestingly, this behavior is at odds with that of rock-paper-scissor games on networks, where heterogeneity shift dynamics towards stable fixed points (see e.g. Ref.~\cite{nagatani2018heterogeneous}).}, \(b^*\) is unique, and iterating \(s\) produces a finite directed cycle: after a transient, the system traverses the same ordered sequence of dominant strains indefinitely (Fig.~\ref{fig:5}). In the limit of a large number of strains, or when trajectories are truncated by host recovery\footnote{Indeed, we have assumed in our evo-SIS model that \(\eta\), which in units of recovery rates represents the typical number of macroscopic mutations occurring during each infection period, is much smaller than one.}, such loops will not close within observation time, yielding exploration of strain space compatible with a short-lived RW. Instead, if \(\mathbf m\) is symmetric --signifying that 1-step mutations are reversible-- recurrent sequences are observed within non-periodic dynamics.

In either the symmetric and non-symmetric cases, we approximate the sequence of dominant strains' identities with a jump process over a \emph{macroscopic mutation network} whose adjacency matrix \(\mathbf Q\) encodes the transition probabilities between dominant strains. In the deterministic limit these probabilities reduce to \(\delta_{b,s(a)}\), but under weak stochastic perturbations --e.g. due to finite pathogen population-- the successor of a given dominant strain becomes random. A natural ansatz for the corresponding transition rates, for \(a \neq b\), is
\begin{equation}
\eta Q_{ba} \approx m_{ba}m_{bb}\left(1-e^{-\kappa/\tau}\right).
\end{equation}
This fixes a correspondence with the host-level mutation rate \(\eta\) as defined in the main text: \(\eta \approx \varepsilon \left(1-e^{-\kappa/\tau}\right)\). In the evo-SIS model, for simplicity, we consider an unweighted mutation matrix, i.e. \(Q_{ab} \to \Theta(Q_{ab} - q_\text{thr})\) for the unit step function \(\Theta(\cdot)\) and a threshold value \(q_\text{thr} > 0\). Furthermore, before invoking additional effects such as multiple immune memories, we consider \(\mathbf Q\) to be symmetric, again for the reason that few (typically one or none at all) macroscopic mutation steps can ever be observed per host under the hypothesis of slow mutations, hence preventing a direct discrimination between directed and undirected RW. The macroscopic Laplacian \(\mathbf L_Q = \mathbf D - \mathbf Q\), with \( \mathbf D\) being the degree (diagonal) matrix of \(\mathbf Q\), governs an effective RW dynamics of dominant strain identities,
\begin{equation}
    \dot{\mathbf p}(t)=-\eta \mathbf L_Q \mathbf p(t),
\end{equation}
where \(p_a(t)\) is the probability that strain \(a\) is the current dominant strain. Of course, at this level of description, we are not tracking the absolute size of the pathogen population, and drawing conclusions on pathogen extinction times, giving a direct connection with the recovery of hosts, is out of scope. That might be achieved by an \textit{ab initio} population dynamics model for the pathogen: as seen in the main text for the evo-SIS framework, the dynamics of relative prevalences are then generally coupled to population size.

\section{Emergence of a directed mutation network from multiple immune memories}
\label{app:directed_mutations}
The evo-SIS model~\eqref{eq:model_full} incorporates cross-immunity only at the level of infection events, where a susceptible host is protected from strains that are antigenically similar to the latest encountered strain. However, immunization is likely to affect the chances of mutations producing viable strains within each host, thus impeding the re-emergence of strains for which some immune memory is present. Since our model does not track this aspect, mutations explore the full undirected and unweighted network available to each host, regardless of the host's immune memory, whose information is immediately lost upon infection. As a consequence, dynamics drive the system towards a steady state where all strains are present (if the mutation matrix \(\mathbf Q\) has one connected component), predicting high strain diversity, which is typically not observed in real epidemics. As noted in the main text, this discrepancy can be overcome by appropriately comparing empirical patterns to transient dynamics, where the phenomenology displayed by our framework comprises these realistic patterns. However, other effects might further constrain the framework's prediction and lead to greater realism.

A comprehensive approach should take into account with more nuance the within-host immunity/mutation interplay \textit{ab initio}; we leave this to future works. Here, we instead consider solely how a directed mutation matrix can emerge from multiple immune memories, thus constraining the phenomenology of our framework; this also clarifies how one should interpret this structure consistently on both modeling and epistemic levels.

The starting point to do so is the introduction of \(m\) immune memories, which we incorporate into the mutation process. To illustrate the argument, we consider mutations at the individual level, and decrease the rates of mutation towards immune-memorized strains. Specifically, suppose that a susceptible individual has encountered the strains \(a_1,\dots,a_m\) and is now infected with strain \(b\). We rescale the out-mutation process as

\begin{equation}
    \sum_b Q_{ab} I_b \to \sum_b \overbrace{\prod_{a = a_1}^{a_m}\left(1 - \eta^*_{ab}\right)}^{\equiv \Delta_{ab}^{(m)}} Q_{ab} I_b,
\end{equation}

where \(\eta^*_{ab}\) is an impeding factor ranging between 0 (immune memory have no effects) and 1 (immune memory prevents re-emergence of previously encountered strains); we assume for simplicity that \(\eta^*_{ab} \equiv 1\). Note that \( \Delta_{ab}^{(m)}\) is not a Markov process, as it depends on the strain chain \(a_1,\dots,a_m\) which is determined by the past states of the individual considered. If the mutation rate is slow with respect to the rate of infection, most individuals will typically carry the same one strain, and only that strain, during a given outbreak. As a consequence, \(\Delta_{ab}^{(m)}\) can be approximately considered homogeneous across the host population. By suppressing both infection and mutation events, strains which have emerged in the outbreak are thus prevented from re-emerging. Successive repetitions of this process lead to chained exclusions reflecting the order in which strains emerged. Moreover, even with a finite number \(m\) of immune memories, as the epidemic progresses, no ``backward path'' can reach previous strains, which are separated from the currently circulating strains by a halo of immune suppression in the antigenic space left in the wake of the evolving pathogen. For these reasons, we can consider the limit 

\begin{equation}
    \Delta_{ab} = \lim_{m \to +\infty}\Delta_{ab}^{(m)}.
\end{equation}

Here, \(\Delta_{ab}\) acts as a ``causal screen'' by preventing more recently emerged strains from back-mutating into already suppressed ones: by naming \(t_a\) and \(t_b\) the emergence time of strains \(a\) and \(b\) respectively, we can write \(\Delta_{ab} \sim \Theta\left(t_a - t_b\right)\), where \(\Theta\left(\cdot\right)\) is the unit step function centered at 0; in a more nuanced choice, recognizing the temporal intermingling of different strains, one can utilize a sigmoid function with the same asymptotic properties, allowing temporally close strains to mutate into each other without fine tuning of emergence times.
The ``screened'' mutation matrix \(\mathbf Q^{\Delta}\), with elements \(Q^{\Delta}_{ab} = \Delta_{ab}Q_{ab}\), assumes a strictly triangular form with  a ladder-like structure dictated by the partial ordering of the emerging times of observed strains; in the case of completely ordered emergence times, \(\mathbf Q^{\Delta}\) assumes exactly (i.e. with a clean diagonal demarcation) a triangular form. In summary, under assumptions of rare mutations and multiple immune memories, the mutation network becomes directed, with the adjacency matrix \(Q\) assuming a strictly triangular form in the limiting case of complete immune-driven suppression of strain re-emergence. We remark that some effects might counter the phenomenon just described: for example, rare mutation channels and enhanced antigenic evasion for some strains, might lead to the re-emergence of past strains. For pathogens where these effects are strong enough, empirical patterns of strain succession should reflect this clearly; for pathogens such as flu and COVID-19, this doesn't appear to be the case.

The argument just presented for within-host immune effects applies to between-hosts (i.e. infection) events as well, and the cross-infectivity matrix \(\mathbf K\) should be replaced by the screened matrix \(K^{\Delta}_{ab} = \Delta_{ba}K_{ab}\). Note the reversal in the indices of the screen: where, before, only older strains could mutate in newer ones, now on the contrary only newer strains can overcome immune protection deriving from the memory of older strains.

From the preceding argument, it is clear that utilizing a directed mutation network implies additional hypotheses beyond the one assumed by our evo-SIS framework. First, the order of emergence of strains is assumed to be known a priori. Second, and related to the previous point, the mechanism leading to the suppression of past strains --consisting in hosts retaining memory of multiple encountered strains-- is not explicitly included in the model, and \textit{cannot} be included except by means of a considerable increase in mathematical and numerical complexity. It is not clear nor obvious that a mixed approach, using a directed mutation network with our single-memory model, is consistent in full generality. Hence, in the main text, a directed mutation network is momentarily assumed only in order to carry out a consistency check, and to show that this structure --if emerging for \textit{any} reason-- leads to dynamical patterns which closely align with our empirical expectations. The emergence itself is entirely beyond the scope of the present framework. We note that a model addressing these issues at the microscopic level, considering within host pathogen populations, would also generalize the ``dominant-strain-only'' simplified approach we adopted in this work and  discussed in Appendix~\ref{app:RM_within_host}.

\section{Diffusion distance and cross-immunity}\label{app:diffusion_distance}

Consider an unweighted undirected network of size \(N\) with adjacency matrix \(\mathbf A\). One might wish to define distances between pairs of nodes \((i,j)\). Some classical constructions do so by counting paths and weighting them according with their length. For example, in Ref.~\cite{estrada2008communicability}, a factorial weighting is realized through the exponential matrix \(e^A\), which could be appropriate in assessing distances according to some contact dynamics. More recent approaches seek instead to build a notion of distance by exploiting the well-understood properties of random walks (RW)~\cite{de2017diffusion}. In this framework, the problem of quantifying the distance between two nodes is recast in terms of how two random walkers, starting at different nodes, explore the network over a time interval whose length represents a scale parameter.

\vspace{0.5cm}
\emph{Unweighted and degree-weighted Laplacian.} Considering continuous time RW, and writing the probability of the walker being at node \(i\) at time \(t\) as the column vector \(\mathbf p(t) = \left(p_1(t),\dots,p_N(t)\right)^T\), the master equation for time evolution is

\begin{equation}
    \dot{\mathbf{p}} = - L \mathbf{p}
\end{equation}

where \(\mathbf L =\mathbf  D - \mathbf A\), with \(\mathbf D = \text{diag}\left(d_1,\dots,d_n\right)\) being the degree matrix. Considering a node \(\bar l\) with zero degree, it holds \(L_{j\bar l} = L_{\bar l j} = 0 \ \forall j\). In all cases, it holds \(\sum_i L_{ij} = 0\), which is equivalent to conservation of probability.

As is well known, the equilibrium solution of this RW is the homogeneous state\footnote{This assumes the network has one connected component; if not, the following applies to each component individually, according to initial conditions.} \(p_i = 1/N \ \forall i \), which solves the detailed balance equation \(L_{ij}p_j = L_{ji}p_i\). At the physical level, this signifies that higher degree nodes are more likely to be reached, but are also vacated quickly: the rate of each step of the random walker is the same for all links in the network, hence the holding time of each node is inversely proportional to its degree. This is an appropriate modeling choice when the transitions described by links are equivalent and independent; at the same time, structural details of the network are lost in the long time limit, as a walker resides for longer times in less connected nodes.

Instead, one could require that exit rates (and hence holding times) be homogeneous across nodes: this is achieved by considering the degree-weighted Laplacian \(\tilde{\mathbf L} = \mathbf L \mathbf D^{-1}\). In the case that no node is isolated, \(\mathbf  D^{-1} = \text{diag}\left(1/d_1,\dots,1/d_n\right)\) is well-defined and we can write \(\tilde{\mathbf L} = \mathbf I - \mathbf A \mathbf D^{-1}\), with \(\mathbf I\) being the \(N \times N\) identity matrix. However, if some nodes are isolated, this factorization does not hold outright: for such a node \(\bar l\), \(\tilde L_{i\bar l} = \delta_{i \bar l} - A_{i \bar l} / d_{\bar l}\). The division by zero in the second term can be hand-waved as \( A_{i \bar l} = 0\), but then \(\sum_i \tilde L_{i\bar l} = 1 \neq 0\). The correctly weighted Laplacian can be obtained by the following regularization procedure. Define first the regularized degree, \(d_i^{(\epsilon)} = \max\left\{\epsilon, d_i\right\}\) with \(\epsilon \in (0,1)\); accordingly, denote the regularized degree matrix as \(\mathbf D^{(\epsilon)}\). Then, the decomposition \(\mathbf L = \tilde{\mathbf L}^{(\epsilon)}{\mathbf D^{(\epsilon)}}\) is always valid, and both objects on the r.h.s. are well defined for positive \(\epsilon\). In particular,

\begin{equation}
     \tilde L^{(\epsilon)}_{ij}=L_{ij}/d_j^{(\epsilon)} =
        (d_j \delta_{ij} - A_{ij})/d_j^{(\epsilon)}.
\end{equation}

Note that \(L_{ii} = 0\) if \(d_i = 0\) for \(\epsilon > 0\). The limit \(\epsilon \to 0\) can then be taken safely: \(\tilde{\mathbf L} = \lim_{\epsilon \to 0}\tilde{\mathbf L}^{(\epsilon)}\) is a degree-normalized Laplacian whose rows and columns associated with isolated nodes are filled with zeros.

\vspace{0.5cm}
\emph{Diffusion distance.} As noted in the previous paragraph, choosing an unweighted or a degree-weighted Laplacian brings out different features in RWs, according to the different physical requirements imposed on the systems' rates. For the weighted Laplacian, the instantaneous RW distribution \(\mathbf{p}(t)\) tends to better reflect the structural properties of the network, as explored by the RW up to time \(t\) starting from a given initial condition. In defining a node-node distance through RWs, therefore, choosing the degree-normalized Laplacian over the unweighted one can better quantify how the connectivity structure makes two nodes close or far apart. Hence, we consider the diffusion distance defined by

\begin{equation}\label{eq:diff_distance}
    d_{ij}^{2}(t) = ||\mathbf{p}^{(i)}(t) - \mathbf{p}^{(j)}(t)||^2 \ ,
\end{equation}

where \(\mathbf{p}^{(i)}(t)\) is the probability vector of the RW at time \(t \geq 0 \), given that at initial time \(t=0\) its position is deterministically confined at node \(i\): \(\left(p^{(i)}(t=0)\right)_j = \delta_{ij}\). 

Some considerations can give a qualitative behavior of the diffusion distance.

(i) If the network has one connected component, then \(d_{ij} \to 0\) for long times, for all pairs \(i,j\), irrespectively of other properties of the network. 

(ii) The long time behavior for a homogeneous network with two connected components can be approximated as follow. Suppose \(i\) belongs to a component of size \(S\), and \(j\) to another component of size \(N-S\). Then, \(\mathbf{p}^{(i)} \approx \left(1/S,\dots,1/S,0,\dots,0\right)\) and  \(\mathbf{p}^{(j)} \approx \left(0,\dots,0,1/(N-S),\dots,1/(N-S)\right)\) and, since \(\mathbf{p}^{(i)} \cdot \mathbf{p}^{(j)} = 0\), he have \(d_{ij}^2 = \frac{1}{Nx(1-x)}\), where \(x = S/N\). It immediately follows that the distance is maximized for \(x \approx 0,1\), where it is equal to \(N/(N-1)\). Hence, even an isolated node has finite (and close to 1) distance from all other nodes in the long time limit; at a qualitative level, this remains true for heterogeneous networks. 

\begin{figure}
    \centering
    \includegraphics[width=\columnwidth]{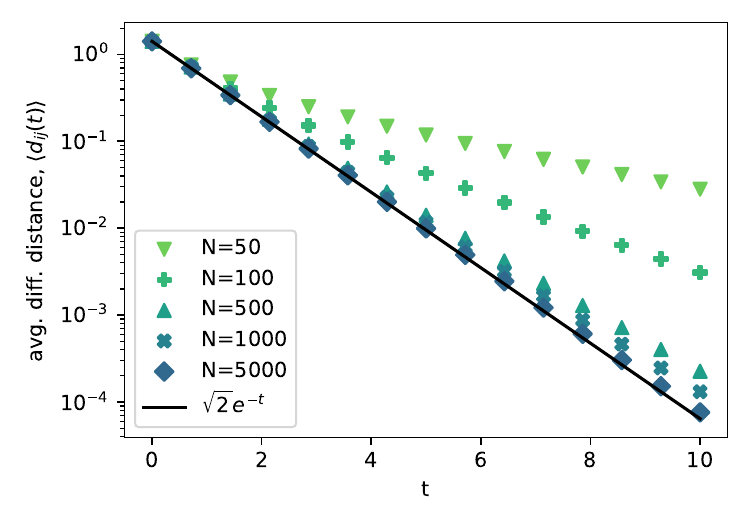} 
    \caption{Diffusion distance at different times \(t\) for a dense Erdos-Renyi network with \(p_\text{conn}=0.1\), and different network sizes \(N\): the theoretical result (solid lines) is approached numerically (markers) as \(N \to +\infty\). Each numerical result is the distance averaged across the network.}
    \label{fig:6}
\end{figure}

(iii) For a dense Erdos-Renyi network, in the limit of large \(N\), one finds that for \(i \neq j\) it holds\footnote{The prefactor \(\sqrt{2}\), which comes from the definition~\eqref{eq:diff_distance}, is irrelevant; it is sufficient to take it into account when dealing e.g. with the epidemic threshold, and in general can be absorbed in \(\beta\).}
\begin{equation}\label{eq:SI_diffdist_limit}
    d_{ij}(t) = \sqrt{2}e^{-t} \ , 
\end{equation}

independently of the connectivity \(p_\text{conn}\) of the network; see Fig.~\ref{fig:6}. In this limit, one does not find isolated nodes in randomly generated networks, but upon forcibly disconnecting node \(\bar l\) (i.e. setting \(A_{i\bar l} = 0 = A_{\bar l i} \ \forall i\)) it is immediate to find

\begin{equation}\label{eq:SI_diffdist_limit_isolated}
    d_{i\bar l}(t) = \left(1+e^{-2t}\right)^{1/2} \ ; 
\end{equation}

we use this result --which is independent from \(p_\text{conn}\)-- to assess the (invasion) epidemic threshold in the main text.

\vspace{0.5cm}
\emph{Modeling mutations and antigenic distance through diffusion.} In this work, we choose to represent mutations as diffusion on an unweighted Laplacian, whereas diffusion distance that describes phenotypic --or, more specifically, antigenic-- similarity, \(\mathbf d_\rho(Q) \equiv \{d_{ij}(\rho)\}_{ij}\) utilizes the degree-weighted Laplacian. 

The first choice follows from our interpretation of host-level mutations given in Appendix~\ref{app:RM_within_host}. There is no particular reason to assume that the (average) holding time of each strain is the same. If a strain has many possible mutations at its disposal, then that strain is less stable and will mutate more frequently, as reflected in a heterogeneous degree distribution for the effective mutation matrix \(\mathbf Q\). Moreover, if diffusion drives towards an homogeneous strain composition --a state in which effective diversity is maximized-- any unevenness in the composition can be attributed to strain selection, resulting from competition  mediated by the host population antigenic profile.

The second choice is instead intended to implement the following assumption: two strains which are connected by many redundant short mutation paths are antigenically more similar than two strains which are only connected through longer mutation chains, and thus more easily induce reciprocal cross-immunization. The fact that a highly connected strain mutates more often does not enter this consideration, as it is strain identity --and not kinetics-- that dictates similarity. Hence, in our framework, time elapsed in the RW is replaced with a parameter, \(\rho\), which controls the structural scale at which strains are considered similar.

We note that, in considering directed mutation structures (which emerge as explained in Appendix~\ref{app:directed_mutations}), the cross-infectivity matrix \(\mathbf K\) has to be obtained from \(\mathbf Q\) before applying the causal mask; only after this, the same mask (now transposed) is to be applied to \(\mathbf K\) itself.

\section{Systematic expansion of evo-SIS dynamics under time scale separation}\label{app:time_scale_separation}

Usually, the simplification of dynamics involving more than one class of variables is carried out rather naively under the hypothesis separation of time scales, by assuming time-stationarity of the variables which one wishes to integrate out of the system in a so-called quasi-stationary approximation. This is a valid approach when these variables are \textit{fast}, and their values are effectively constrained to be a function of the remaining \textit{slow} variables.

Here, the fast variables to be integrated out are the susceptible compartments, \(S_{xa}\). Instead of directly assuming \(\dot S_{xa} = 0\), we follow a more systematic procedure, recovering at leading order (LO) the quasi-stationary result, but also explicitly showing how higher-order terms arise. Moreover, this calculation naturally allows for the necessary considerations regarding the constrained space of the effective dynamics.

We remark that, in the following, we shall occasionally use powers of $\beta$ as a useful device to track terms at different orders of the expansion; however, the conditions which separates different orders pertains the Forces of Infection (FoI), as defined and explained in the following section.

\subsection{Leading-order effective dynamics}\label{app:time_scale_separation_1}

Starting from eqs.~\eqref{eq:model_full}, we integrate the dynamics of \(S_{xa}\) in terms of the full set of infected compartments, \(\left\{I_{xa}\right\}_{xa}\), which involves no additional assumption or approximation. This procedure allows for closure of dynamics for \(I_{xa}\) as integro-differential equations. First, susceptibles' state variables are completely determined in terms of infected state variables, and read
\begin{multline}\label{eq:system_full_integrodiff}
    S_{xa}(t)\equiv S\left[I\right]_{xa}(t) =S_{xa}(0)e^{-\int_{0}^{t}\lambda_{xa}(\tau)\,d\tau} \\+\int_{0}^{t} e^{-\int_{t-u}^{t}\lambda_{xa}(\tau)\,d\tau}\,I_{xa}(t-u)\,du ,
\end{multline}
where we defined the FoIs
\begin{equation}\label{eq:exposure_rate}
\lambda_{xa}(t) \equiv \beta\sum_{y=1}^N\sum_{b=1}^M A_{xy}\,K_{ab}\,I_{yb}(t) \ .
\end{equation}

In eq.~\eqref{eq:system_full_integrodiff}, the first term simply relays how quickly initial conditions are forgotten with time; we will neglect this in the following. The second term introduces memory of previous infection states. By plugging eq.~\eqref{eq:system_full_integrodiff} into eq.~\eqref{eq:model_full1}, we obtain a delay differential equation for strains' prevalence, equivalent to the full dynamics~\eqref{eq:model_full}, where the current state of susceptibles is completely encoded in the past states of infected individuals. We can rewrite the density of each infect class, at each node, as \(I_{xa} = I p_{xa}\), where \(I = \sum_{xa}I_{xa}\) is the total incidence. Note that this allows us to write --using a compact matrix notation-- \(\lambda_{xa} = \beta I(\mathbf A p \mathbf K^T)_{xa} \equiv \beta I \hat \lambda_{xa}\), factorizing the epidemic size. We assume that \(\hat \lambda_{xa} = \mathcal{O}\left(1\right) = I\); from these two equalities, consistently, it will follow in the expansion that \(\dot{\hat{\lambda}}_{xa} = \mathcal{O}\left(1\right)\) while \(\dot{I} = \mathcal{O}\left(\beta^{-1}\right)\). The original system~\eqref{eq:model_full} is readily rewritten (all time arguments outside the memory terms are understood to be equal to \(t\)) as

\begin{widetext}
\begin{align}\label{eq:system_full_integrodiff_epidemicsizeexplicit}
        \nonumber \dot{p}_{xa} &= +\beta \sum_y A_{xy}p_{ya}\sum_b K_{ba}S_{xb} - p_{xa} - \eta \sum_b \left(L_Q\right)_{ab} p_{xb} \\ 
        &\equiv \sum_y\underbrace{\left[ A_{xy}R_{x}^a[p|\beta I]  - \delta_{xy}\left( \sum_{x',y',a'}A_{x'y'}R_{x'}^{a'}\left[p|\beta I\right]p_{y'a'}\right)\right]}_{\phi_{xy}^a[p|\beta I] - \delta_{xy}\bar\phi[p|\beta I]}p_{ya} - \eta \sum_b \left(L_Q\right)_{ab} p_{xb} \ ,
\end{align}
\end{widetext}

where we have defined

\begin{multline}\label{eq:R}
    R_{x}^a[p|\beta I] = \\\sum_b K_{ba}\int_{0}^{t} e^{-\int_{t-u}^{t}\beta I(\tau)\hat\lambda_{xb}(\tau)d\tau}\,\beta  I(t-u)p_{xb}(t-u)du ;
\end{multline}

the last term in the square brackets of eq.~\eqref{eq:system_full_integrodiff_epidemicsizeexplicit}, coming from the normalization requirement on the relative prevalences, is the global mean fitness. Note that \(I(t)\) has to be evolved through a separate differential equation, but can be treated with the same formalism, since

\begin{equation}\label{eq:total_prevalence_dynamics_integrodifferential}
    \dot I = \left[ \sum_{x,y,a}A_{xy}R_{x}^a\left[p|\beta I\right]p_{ya} -1\right]I \ .
\end{equation}

Eqs.~\eqref{eq:system_full_integrodiff_epidemicsizeexplicit} and ~\eqref{eq:total_prevalence_dynamics_integrodifferential}, taken together, are equivalent \edit{(up to the neglected transient, which is forgotten exponentially fast)} to eqs.~\eqref{eq:model_full}. The factor \(R^a_{x}\) shows how the epidemic size \(\beta I(t)\) dictates the rates of contagion in eq.~\eqref{eq:system_full_integrodiff_epidemicsizeexplicit}. Conversely, the exponential weighting is such that when \(\beta I(t)\) is large, memory of previous states fades quickly. Owing to this fact, despite this equation not being solvable in full generality, in this regime of large epidemic size we can make progress and gain some insight on the dynamics of strain diversity.

To this end, we expand the integral definition of \(R_{x}^a\) as reported in  eq.~\eqref{eq:R} \edit{with respect to inverse powers of \(\beta I\)}. There, by our assumptions on the order of magnitude of \(I\) and \(\hat\lambda_{xa}\), the exponential factor is non-vanishing only for \(u \ll t\). We thus expand \(\int_{t-u}^{t}I(\tau)\hat\lambda_{xa}(\tau)\,d\tau = I(t)\hat\lambda_{xa}(t)\,u + \mathcal{O}(u^2)\) and \(I(t-u)p_{xb}(t-u) = I(t)p_{xb}(t) - \frac{d}{dt}\left(Ip_{xb}\right)(t)\,u + \mathcal{O}(u^2)\). Dropping subleading terms, and extending the upper boundary of the outer integral from \(\tau = t\) to \(\tau = +\infty\) with exponentially vanishing error, we obtain

\begin{align*}
    R_x^a &= \sum_b K_{ba}\int_{0}^{t} e^{-\int_{t-u}^{t}\beta I(\tau)\hat\lambda_{xb}(\tau)\,d\tau}\,\beta I_{xa}(t-u)\,du \\ &\approx\sum_b K_{ba} \int_{0}^{+\infty} e^{-\beta I(t)\hat\lambda_{xb}(t)u}\beta I(t)\left(p_{xb}(t)-\dot{p}_{xb}(t)u\right)du \\ &=  \sum_b K_{ba} \left[\frac{p_{xa}(t)}{\hat\lambda_{xa}(t)} - \frac{1}{\beta I(t)}\frac{\dot{p}_{xa}(t)}{\left(\hat{\lambda}_{xa}(t)\right)^2}\right]\ .
\end{align*}

Here, the LO component of \(R_x^a\) already emerges clearly, however there is no guarantee that the NLO is complete. 

To put the previous procedure on a more \edit{systematic} footing, we note that the same expansion can be easily performed by rewriting eq.~\eqref{eq:system_full_integrodiff} as the formal inversion of the differential operator which defines susceptible dynamics:

\begin{multline}\label{eq:system_full_resolvent_series}
    R_{x}^a[p|\beta I](t) \equiv \beta \sum_b K_{ba}S_{xb}(t) \\=  \beta \sum_b K_{ba}\left(\frac{d}{dt} + \beta I\hat\lambda_{xb}(t) \right)^{-1}I_{xb}(t) \\= \sum_b \frac{K_{ba}}{I(t)\hat{\lambda}_{xb}(t)}\sum_{s=0}^{+\infty}\left(-\frac{d}{dt}\frac{1}{\beta I(t)\hat{\lambda}_{xb}(t)}\right)^s \left(I(t) p_{xb}(t)\right)  \ .    
\end{multline}

We remark that, in the last passage of eq.~\eqref{eq:system_full_resolvent_series}, the correct choice of order in which to place the time derivative and the time-dependent factor \(1/I\hat\lambda_{xb}\) can be easily determined from a comparison with the integral expansion which we carried out previously. At each order, evaluating the time derivatives which appear in the expansion require a closure scheme: we simply insert the dynamics at the previous order. The resolved series thus obtained\footnote{\edit{The convergence (or lack thereof) of the resolvent series could be proved rigorously by calculating the spectral gap of the operator of which the series represents the inverse. The spectral gap quantifies the time scale separation between fast relaxation modes and secular variations. The larger is the gap, the more the fundamental state (i.e. the slow manifold) is robust to perturbations. Conversely, as the gap decreases, the system displays correlations over long time scales, and fast modes become relevant. Given the discussion carried out in the appendix, a naive estimate of the gap is \(\inf_{t,a,x}\beta I(t)\hat{\lambda}_{xa}(t)\).}} is equivalent to the one coming from the expansion of the integral above, and it constitutes a useful device to keep track of all terms of the same order. On a formal level, then, the adiabatic expansion is tantamount to considering the series

\begin{equation}\label{eq:R_series}
    R_{xy}^a[p|\beta I](t) = \sum_{s=0}^{+\infty}\left(\beta I(t\right))^{-s} \  R_{x}^{a(s)}\left(p(t)\right) \ ,
\end{equation}

in which the first scale-independent coefficient is

\begin{equation}\label{eq:R_LO}
    R_x^a \equiv R_x^{a(0)} = \sum_b \frac{K_{ba} p_{xb}}{\sum_{z,c}A_{xz}K_{bc}p_{zc}} \ .
\end{equation}

Hence, by plugging~\eqref{eq:R_LO} into eq.~\eqref{eq:system_full_integrodiff} we obtain the LO dynamics

\begin{multline}\label{eq:model_expansion_1st}
    \dot{p}_{xa} = \sum_y \overbrace{\left[A_{xy}\sum_b \frac{K_{ba} p_{xb}}{\sum_{z,c}A_{xz}K_{bc}p_{zc}} - \delta_{xy}\right]}^{\equiv \phi_{xy}^a}p_{ya} \\ - \eta \sum_b \left(L_Q\right)_{ab}p_{xb} \ ,
\end{multline}

which is the RM equation~\eqref{eq:model_RM} in the main text. In this general case of heterogeneous metapopulation structure, the fitness term mixes relative prevalences at different nodes in the metapopulation network, which is further discussed in the main text. Note that the mean fitness is one. Relatedly, at LO, the model describes a trivial evolution of epidemic size: the total prevalence \(I(t)\) is constant over time, as

\begin{multline}
    \dot{I}/I = \sum_{x,y,a}A_{xy}\sum_b \frac{K_{ba} p_{xb} p_{ya}}{\sum_{z,c}A_{xz}K_{bc}p_{zc}}-1 \\
    = \sum_{x,b} \underbrace{\frac{\sum_{y,a}A_{xy}K_{ba}p_{ya}}{\sum_{z,c}A_{xz}K_{bc}p_{zc}}}_{=1}p_{xb} - 1 \equiv 0 \ .
\end{multline}

Carrying on the expansion at next-to-leading-order (NLO) of the adiabatic series,
one finds

\begin{equation}\label{eq_R_NLO}
    R_x^{a(1)} = \sum_b K_{ba}\left[-\frac{\dot p_{xb}}{\left(\hat\lambda_{xb}\right)^2} + \frac{\dot{\hat{\lambda}}_{xb}p_{xb}}{\left(\hat\lambda_{xb}\right)^3} \right] \ ,
\end{equation}

where \(\dot p_{xb}\) and \(\dot{\hat{\lambda}}_{xb}\) are calculated at LO, which lead to eqs.~\eqref{eq:model_homogeneous_2nd_order} in the homogeneous case.

\subsection{Manifold of effective dynamics}\label{app:time_scale_separation_3}

In considering the full dynamics~\eqref{eq:model_full}, the constraint of local mass (or probability) conservation, i.e. \(\frac{d}{dt}\sum_{a}\left(S_{xa} + I_{xa}\right) \equiv 0 \ \forall x\), is automatically satisfied. However, each order of the adiabatic expansion operated by means of the series~\eqref{eq:system_full_resolvent_series} equates \(S_{xa}\) with a sum of differential operators acting on the full set \(\{I_{xa}\}_{xa}\). As obtaining the effective (adiabatic) dynamics at order \(q\) consists in truncating the resolvent series~\eqref{eq:system_full_resolvent_series} at \(s=q\), the order-\(q\) manifold \(\mathcal M_q \) can be written compactly as

\begin{widetext}
    \begin{equation}\label{eq:effective_dynamics_manifold}
    \mathcal M_q = \left\{ I_{xa} \left| \frac{d}{dt} \sum_a \left[1 + \frac{1}{\lambda_{xa}(t) }\sum_{s=0}^{q}\left(-\frac{d}{dt}\frac{1}{\lambda_{xa}(t) }\right)^s\right]I_{xa}(t) = 0\right. \right\}
    \end{equation}
\end{widetext}

It is understood that time derivatives require substitution of the equivalent terms obtained from the equations of \(I_{xa}\).

For example, at LO (i.e. \(q=0\)) it has to hold at each node \(x\)

\begin{equation}
   \sum_a I_{xa} \left(1 + \lambda_{xa}^{-1}\right) = \text{const.} \ ,
\end{equation}

where the constant r.h.s is dictated by the choice of normalization. Initial conditions also have to underlie this constraint, to have a consistent temporal evolution, failing which the full dynamics and its approximants will in general describe different trajectories. \edit{This is also necessary to compensate the fact that we have neglected the transient term in eq.~\eqref{eq:system_full_integrodiff}.}

\section{Computational details}\label{app:computational_details}

All the code to perform numerical simulations and the analysis of results was written in \texttt{Python}, and optimized with \texttt{Numba}. Numerical integration was carried out through a fourth-order Runge-Kutta scheme; evaluation times were predetermined and used logarithmically scaled time steps in the initial phase of integration (i.e. until time \(t\) such that \(t/t_\text{max} \leq 10\)), then switching to uniform time steps. This ensured smooth transients and numerical stability.

In comparing approximant dynamics with the unapproxximated dynamics in eq.~\eqref{eq:model_homMF}, initial conditions satisfying the quasi-stationary constrain~\eqref{eq:effective_dynamics_manifold} are determined as follow. An initial seed \(\{I_a^*\}_a\) is defined for absolute prevalences, with \(\sum_a I_a^* = 1\). Then, we seek the scale factor \(s\) such that \(I_a = s I_a^*\) satisfies the LO constraint, \[\sum_a I_a\left(1+
\frac{1}{\beta(\mathbf KI)_a}\right) = 1\ ,\] which can be solved for \(s\) finding \[s = 1 - \frac{1}{\beta}\sum_a \frac{I_a^*}{(\mathbf K I^*)_a}\ .\] At NLO, the same procedure leads to a quadratic equation for \(s\).

The sampling of networks has been carried out with commonly used algorithms; we used random number generators given by the \texttt{NumPy} library.

\subsection{Details on Main Text Figures}

\noindent \textbf{Fig.~\ref{fig:1}:}  Cities and pathogen sprites have been obtained at~\url{https://www.svgrepo.com/}.
\newline \noindent \textbf{Fig.~\ref{fig:2}:} \textbf{(IT\(\infty\))},\textbf{(ET\(\infty\)):} \(M=100\). For each fixed set of parametres \(\beta,\rho,\eta\), we carried out 200 simulations, each time resampling \(\mathbf{Q}\) from an Erdos-Renyi ensamble with \(p_\text{conn}=0.1\). At the end of simulations, \(t_\text{max}=10^3\), the system reaches the steady state: the subfigures report final epidemic size \(I = \sum_a I_a\) and evenness as defined in eq.~\eqref{eq:evenness} calculated over the final state.|| \textbf{(PE),(DT),(NO):} \(M=200\), \(\eta=10^{-6}\). \(\beta\) has been chosen to match a fraction of the critical value \(\beta_\text{crit}^*\) as described in the caption of the Figure. Each simulation run from \(t_0=0\) to \(t_\text{fin} = 5000\); 500 simulations per point have been used to compute average values of the quantities involved, with \(Q\) (exponential random geomtric network, \(p_0 = 0.9\) and broadness parameter \(=0.02\)) being resampled at each simulation. The probability of reaching an endemic steady state has been defined as \(\text{prob}\left(I(t_\text{fin})\right) > 10^{-3}\); results depend weakly on this threshold.
\newline \noindent \textbf{Fig. 3:} We fit the models prediction to the empirical observables --mean strain dominance time and peak number of infected-- by gridsearch over likely sets of parameters. This selected set of parameters is then used as the initial point for a square loss minimization over the full trajectories; however, loss fluctuates due to randomness in \(\textbf Q\) without attaining a well-defined minimum value. Thus, we set \(\rho = 10\), \(\beta/\beta_\text{crit}^* = 1\), \(\eta = 0.01\), according to the result of the initial gridsearch. \(Q\) is taken to be a random geometric network of basic connectivity \(p_0 = 1\) and broadness parameter \(r = 0.04\), which itself has been determined by the same procedure as the model's rates. We note that the determination of said parameters is not unique, due to the roughness of the target function to be minimized, that is square loss between model output and data. Moreover, the goodness of prediciton is conditioned on the sampled mutation matrix \(\mathbf{Q}\) having one connected component. In particular, predicitons fail if the seed strain is located at an isolated node, or at a node belonging to a small connected component.
\newline \noindent \textbf{Fig. 4:} \textbf{(E),(TI):} \(M=50\), \(\rho =1.4\), \(\beta =21\), \(\eta=0.0014\). LO and NLO dynamics are integrated in the form of eqs.~\eqref{eq:model_homogeneous_RM} and~\eqref{eq:model_homogeneous_2nd_order}; similarly for the NNLO dynamics. \(Q\) is an Erdos-Renyi network with connection probability \(p_\text{conn} = 0.15\), identical for all simulations. Initial conditions have been chosen as explained earlier in this Appendix~\ref{app:computational_details}. ||\textbf{(ED)} The spectrum contains the eigenvalues of the `supra-Laplacian' \(\Gamma_{xa}^{yb}\) which has been represented as a matrix by redefinition of labels, i.e. \(\Gamma_{xa}^{yb} \to \Gamma_i^j\).  ||\textbf{(TO)} Data have been smoothed with a Gaussian filter to abate slight fluctuations, which persist even with relatively high statistic (here each point is computed by 100 realizations of the same simulation with resampled mutation networks). Each deme of hosts carries different \(Q^x\) randomly drawn from a set of increasing size \(N_Q\) from the same ensemble, an exponential random geometric network with basic connectivity \(p_\text{conn}^Q = 0.2\) and broadness \(r = 0.05\). Model parameters' and initial conditions are identical in all cases. Single realizations of the mutation network \(Q\) have several connected components, so that the strain space cannot be fully explored within a single host. The probability of reaching an endemic state is here defined as the probability that \(I(t\to+\infty) < I(0) = 10^{-4}\); results depend only quantitatively on this choice.
\\


\bibliography{apssamp}

\end{document}